\documentclass[11pt,a4paper]{article}
\usepackage{cite,mathtools,amsmath,amssymb,color,comment,graphicx}
\usepackage[usenames]{xcolor}
\usepackage[bookmarksopen,colorlinks=true,linkcolor=dark_green,citecolor=dark_red,urlcolor=dark_red,linktocpage=false]{hyperref}
\usepackage[height=22.5cm,width=16.5cm,centering]{geometry}
\usepackage[normalem]{ulem}

\definecolor{dark_blue}{rgb}{0,0,0.6}
\definecolor{dark_green}{rgb}{0,0.4,0}
\definecolor{dark_red}{rgb}{0.6,0,0}
\definecolor{orange}{rgb}{1.,0.67,0.2}

\newcommand{\nn}{\nonumber} 
\newcommand{\be}{\begin{equation}} 
\newcommand{\ee}{\end{equation}} 
\newcommand{\bea}{\begin{eqnarray}} 
\newcommand{\eea}{\end{eqnarray}}

\begin{document}
%%%%%%%%%%%%%%%%%%%%%%%%%%%%%%%%%%%%%%%%%%%%%%%%%%

%%%%%%%%%%%%%%%%%%%%%%%%%%%%%%%%%%%%%%%%%%%%%%%%%%
\begin{titlepage}

\begin{center}

\hfill DESY 21-131 \\

\vskip 2cm

{\fontsize{15pt}{0pt} \bf
Effect of density fluctuations on
}
\vskip 0.5cm
{\fontsize{15pt}{0pt} \bf 
gravitational wave production in first-order phase transitions
}

\vskip 1.2cm

{\Large
Ryusuke Jinno, Thomas Konstandin,
}
\\[0.4cm]
{\Large
Henrique Rubira and Jorinde van de Vis
}

\vskip 0.4cm

\end{center}
\begin{center}

\textsl{Deutsches Elektronen-Synchrotron DESY, Notkestr. 85, 22607 Hamburg, Germany}
\vskip 8pt

\end{center}

\vskip 1.2cm

\begin{abstract}
We study the effect of density perturbations on the process of first-order phase transitions and gravitational wave production in the early Universe. We are mainly interested in how the distribution of nucleated bubbles is affected by fluctuations in the local
temperature.
We find that large-scale density fluctuations ($H_* < k_* < \beta$) result in a larger effective bubble size at the time of collision, enhancing the produced amplitude of gravitational waves.
The amplitude of the density fluctuations necessary for this enhancement is ${\cal P}_\zeta (k_*) \gtrsim (\beta / H_*)^{-2}$, and therefore the gravitational wave signal from first-order phase transitions with relatively large $\beta / H_*$ can be 
significantly enhanced by this mechanism even for fluctuations with moderate amplitudes. 
\end{abstract}

\end{titlepage}

\tableofcontents
\thispagestyle{empty}

\renewcommand{\thepage}{\arabic{page}}
%%%%%%%%%%%%%%%%%%%%%%%%%%%%%%%%%%%%%%%%%%%%%%%%%%

\newpage
\setcounter{page}{1}

%%%%%%%%%%%%%%%%%%%%%%%%%%%%%%%%%%%%%%%%%%%%%%%%%%
\section{Introduction}
\label{sec:Introduction}
%%%%%%%%%%%%%%%%%%%%%%%%%%%%%%%%%%%%%%%%%%%%%%%%%%

The groundbreaking detection of gravitational waves (GWs) by the LIGO collaboration marks a new era in the GW science program \cite{Abbott:2016blz}. A next step in this agenda will be carried out by LISA in the next decade \cite{Seoane:2013qna, amaroseoane2017laser}, which is sensitive to astrophysical sources at lower frequencies. Besides the focus on astrophysical objects, LISA has the potential to observe a stochastic GW background (SGWB) with a peak frequency in the mHz regime. 
Among the different kinds of sources of a SGWB at mHz, a very appealing one is a first-order phase transition 
 with a temperature of around the TeV scale \cite{Coleman:1977py, Linde:1980tt, Steinhardt:1981ct,Witten:1984rs}. A first-order phase transition at that scale is present in many extensions of the Standard Model (SM) and can in principle seed the baryon asymmetry of our universe \cite{Kuzmin:1985mm, Cohen:1993nk, Rubakov:1996vz, Riotto:1999yt,Morrissey:2012db}. For references on the GW science program, and in particular the prospects of LISA, see e.g. \cite{Caprini:2015zlo, Cai:2017cbj, Caprini:2019egz, Barausse:2020rsu, Hindmarsh:2020hop, Bian:2021ini}, and for other proposals on GW detection around the LISA frequency band, see e.g. \cite{Seto:2001qf, TianQin:2015yph, Hu:2017mde, AEDGE:2019nxb, Sesana:2019vho, Badurina:2019hst, Blas:2021mqw}.

A first-order phase transition triggered by a scalar field generates true-vacuum bubbles at disjoint regions in space. Depending on the strength of the transition and the thermodynamic properties of the coupling between the scalar field and the plasma \cite{Espinosa:2010hh,Giese:2020rtr,Giese:2020znk}, those bubbles can expand and eventually collide, sourcing GWs. Simulations have been built to fully understand and parametrize the contributions to the GW spectrum coming from the scalar-field solitons and the plasma sound waves~\cite{Hindmarsh:2013xza,Hindmarsh:2015qta,Hindmarsh:2017gnf,Cutting:2019zws,Jinno:2020eqg}.
These simulations show that for strong enough phase transitions, the GW signal is indeed in reach of experiments such as LISA.

Furthermore, it has been shown that the SGWB spectrum and its anisotropies could be used to probe curvature fluctuations \cite{Bartolo:2019oiq, Domcke:2020xmn}.
A yet unexplored question is how the SGWB sourced by a first-order phase transition could be affected by temperature fluctuations coming from an inflation-like scenario. Whilst the power spectrum of primordial fluctuations is well-measured by CMB observations on very large scales and constrained by primordial black holes on small scales, little is known about it on even smaller scales~\cite{Bringmann:2011ut}. Some feature in the inflationary potential or particle production during inflation may enhance curvature perturbations with high wavenumber \cite{Sasaki:2018dmp,Linde:2012bt}. The presence of fluctuations in the plasma can affect bubble nucleation and generate imprints in the GW spectrum.
We investigate in this work how first-order phase transitions are affected by temperature fluctuations. As we make clear later in the text, the GW spectrum is mostly affected by scales $H_* < k_* < \beta$, where $H_*^{-1}$ is the Hubble radius during nucleation and $\beta^{-1}$ the typical bubble size. Our main result is that these temperature fluctuations will 
increase the average bubble size which leads to an enhancement of the GW signal. This might be quite intuitive for 
static temperature fluctuations: bubbles nucleate mostly in the cold spots which increases the effective bubble size at the time of collision.
We show that this effect also persists when the temperature fluctuations propagate as sound waves.

The organization of the paper is as follows.
In Sec.~\ref{sec:GWs} we review GW production in first-order phase transitions and explain our simulation scheme.
In Sec.~\ref{sec:Idea} we present the central idea of this paper on the enhancement of GWs in the presence of temperature fluctuations.
In Sec.~\ref{sec:Results} we prove this idea with numerical simulations.
Sec.~\ref{sec:DC} is devoted to discussion and conclusions.

%%%%%%%%%%%%%%%%%%%%%%%%%%%%%%%%%%%%%%%%%%%%%%%%%%
\section{Gravitational waves in first-order phase transitions}
\label{sec:GWs}
%%%%%%%%%%%%%%%%%%%%%%%%%%%%%%%%%%%%%%%%%%%%%%%%%%

Our interest in the present paper is the production of the tensor components $h_{ij}$ of the metric 
\begin{align} \label{eq:metric}
ds^2
&=
g_{\mu \nu} dx^\mu dx^\nu 
=
- dt^2 + a^2 (\delta_{ij} + h_{ij}) dx^i dx^j\,.
\end{align}
The time evolution of $h_{ij}$ in transverse-traceless gauge for each Fourier component is given by
\begin{align} \label{eq:heq}
\ddot{h}_{ij} + k^2 h_{ij}
&=
\frac{2}{M_P^2} \Lambda_{ij,kl} T_{kl}\,,
\end{align}
where $T_{ij}$ is the energy-momentum tensor of the system, $M_P = 1 / \sqrt{8 \pi G}$ is the reduced Planck mass, and $\Lambda_{ij,kl} = P_{ik} P_{jl} - P_{ij} P_{kl} / 2$ with $P_{ij} = \delta_{ij} - \hat{k}_i \hat{k}_j$ the projection tensor onto the transverse-traceless components of the tensor.
We neglect the cosmic expansion during the transition.
In typical first-order phase transitions, the energy-momentum tensor $T_{ij}$ results from the fluid dynamics while the 
contribution from the scalar field driving the phase transition is negligible.

In this paper we assume a perfect fluid with a relativistic equation of state
\begin{align}
T_{\mu\nu}
&=
w u_\mu u_\nu + p g_{\mu \nu}\,,
\end{align}
where $w$, $p$, and $u^\mu$ are the enthalpy density, pressure, and fluid four-velocity, respectively.
Given the source term, the GW spectrum is calculated from Weinberg's formula~\cite{Weinberg:1972kfs}
\begin{align}
\Omega_{\rm GW} (q)
&\equiv 
\frac{1}{\rho_{\rm tot}} \frac{d \rho_{\rm GW}}{d \ln q}
=
\frac{q^3}{4 \pi^2 \rho_{\rm tot} M_P^2 V_{\rm sim}}
\int \frac{d\Omega_k}{4\pi} 
\left[ \Lambda_{ij,kl} T_{ij}(q, \vec{k}) T_{kl}^*(q, \vec{k}) \right]_{q = k}\,,
\label{eq:OmegaGW}
\end{align}
where $V_{\rm sim}$ is the simulation volume, $\rho_{\rm tot}$ is the total energy density of the Universe, $q$ and $\vec{k}$ are GW frequency and wave vector, respectively, and $k$ is the norm of $\vec{k}$. 
Moreover, $T_{ij} (q, \vec{k})$ is the Fourier transform of the energy-momentum tensor, and the Fourier transforms in the space and time directions are performed over the finite simulation volume and total duration of the source respectively. 
In first-order phase transitions, compression waves are known to act as a long-lasting source of GWs.
Since the GW spectrum grows linearly in time, we can discuss the growth rate (or power) of the GW spectrum.
After factoring out trivial dependencies, we get
\begin{align}
\Omega_{\rm GW} (q)
&=
\frac{w^2 \; \tau}{4 \pi^2 \rho_{\rm tot} M_P^2 \beta}
\times Q' (q),
\end{align}
where $\tau$ is the lifetime of the sound waves, which typically is determined by the onset of turbulence and bounded from above by $H_*^{-1}$ with $H_*$ being the Hubble parameter around the transition time~\cite{Ellis:2020awk}.
Here, we extracted factors of enthalpy $w$ and the Planck mass in order to make the scaling explicit and the prefactor and $Q'$ dimensionless.
Note that the prime in $Q'$ indicates the ($\beta$-normalized) time derivative, not the derivative with respect to its argument $q$.

From Eq.~(\ref{eq:OmegaGW}), the power of the GW emission $Q'$ becomes
\begin{align}
Q' (q)
&\equiv
\frac{q^3 \beta}{w^2 \; V_{\rm sim} \; T_{\rm sim}}
\int \frac{d\Omega_k}{4\pi} 
\left[ \Lambda_{ij,kl} T_{ij}(q, \vec{k}) T_{kl}^*(q, \vec{k}) \right]_{q = k}\,,
\label{eq:Qpr}
\end{align}
where $T_{\rm sim}$ is the simulation time. Note that the time integration in the Fourier transform of $T_{ij}$ now only runs over the simulation time. Assuming a weak phase transition and a radiation equation of state, the relative factor between $\Omega_{\rm GW}$ and $Q'$ can be rewritten in terms of $H_*$ as
\begin{align}
\frac{\Omega_{\rm GW} (q)}{Q^\prime (q)}
&=
\frac{w^2 \; \tau}{4 \pi^2 \rho_{\rm tot} M_P^2 \beta} \simeq 
\frac{4 \rho_{\rm tot} \; \tau}{9 \pi^2 M_P^2 \beta} 
= 
\frac{4 H_* \tau}{3 \pi^2} \frac{H_*}{\beta} \; ,
\end{align}
meaning $\Omega_{\rm GW} \sim Q' \times (H_* / \beta)$ for sound waves lasting for the entire Hubble time.

Already at this stage, it can be understood how larger bubbles can lead to a potentially larger signal. The factors $\tau$
and $1/\beta$ basically arise from the two-point function of the energy-momentum tensor and parametrize the correlation 
time and length of the process, $R_* \simeq v_w/\beta$ where $R_*$ is the average bubble size and $v_w$ the velocity of the phase transition fronts. Increasing the correlation length will lead to a stronger GW signal.

%%%%%%%%%%%%%%%%%%%%%%%%%%%%%%%%%%%%%%%%%%%%%%%%%%
\section{Impact of temperature fluctuations}
\label{sec:Idea}
%%%%%%%%%%%%%%%%%%%%%%%%%%%%%%%%%%%%%%%%%%%%%%%%%%

One of the key ingredients in the GW production from first-order phase transitions is the bubble nucleation rate.
The spacetime distribution of the nucleation points determines the GW spectrum.
In finite-temperature transitions, the dominant part of the nucleation rate is determined from the three-dimensional bounce action $S_3 / T$
\be
\Gamma
\propto e^{- S_3 / T}. \label{eq:rate}
\ee
Expanding the tunneling action $S_3$ around the typical transition time $t = t_*$ and in temperature fluctuations, $\delta T$, we obtain
\begin{align}
\Gamma
&=
\Gamma_* \exp \left[ \beta (t - t_*) - \frac{\beta}{H_*} \frac{\delta T}{\overline{T}} \right],
\end{align}
with $\Gamma_*$ being the nucleation rate at the typical transition time $t = t_*$, and quantities with a bar indicate background values. 

In order to study how the temperature fluctuations affect the spatial distribution of the bubbles, we implemented an algorithm including temperature fluctuations to nucleate the bubbles (see Appendix~\ref{app:algorithm} for a complete description of the algorithm used). We now specify how we set the initial conditions for the temperature fluctuations and parametrize them in terms of the wavenumber $k_\ast$ and the variance of the fluctuations $\sigma$.

%%%%%%%%%%%%%%%%%%%%%%%%%%%%%%%%%%%%%%%%%%%%%%%%%%
\paragraph{Initial conditions.}
%%%%%%%%%%%%%%%%%%%%%%%%%%%%%%%%%%%%%%%%%%%%%%%%%%

For notational simplicity we define the normalized temperature fluctuation
\begin{align}
\frac{\beta}{H_*} \frac{\delta T}{\overline{T}}
&=
\delta \tilde{T}\,.
\end{align}
Naively one expects that a significant impact on the bubble nucleation distribution 
requires $\delta \tilde{T} \simeq 1$. Typically, the duration of the phase transition
is quite short compared to the Hubble parameter $\beta \gg H_*$ 
and hence significant effects are expected already for small fluctuations 
\be
\beta \gg H_*, \quad \delta T/\overline{T} \ll 1 \,.
\label{eq:Tsmall}
\ee
Throughout the paper we assume that the temperature fluctuations are small enough and do not affect the wall velocity 
which is a good approximation in the light of Eq.~(\ref{eq:Tsmall}).
We assume that the temperature fluctuations are Gaussian and initialize the Fourier modes in 
our simulation with random phases.

We parametrize density fluctuations in terms of their typical scale and amplitude as follows.
The fluctuations have a typical wavenumber $k_*$. In the simulations, the spectrum of fluctuations is a top-hat 
distribution (between $k_*$ and $k_*/2$) with random phases. The amplitudes of the fluctuations are also 
randomized and the total power in the fluctuations is given by
\be
\sigma^2 = \frac{1}{V}\int d^3x \, \delta \tilde{T}(x)^2 \simeq \frac{1}{N^3}\sum_{x_i} \delta \tilde{T}(x_i)^2 \, ,
\ee
which in Fourier space can be expressed as
\be
\sigma^2 = \int \frac{d^3k}{(2\pi)^3} P_{\delta\tilde{T}}(k) \, \simeq \frac{1}{(2\pi N)^3} \sum_{k_i} \delta \tilde{T}(k_i)^2\,,
\ee 
with $N$ being the number of grid points in one dimension and $P_{\delta\tilde{T}}$ being the power spectrum of $\delta \tilde{T}$.

%%%%%%%%%%
\begin{figure}[t]
\centering
\includegraphics[width=0.32\textwidth]{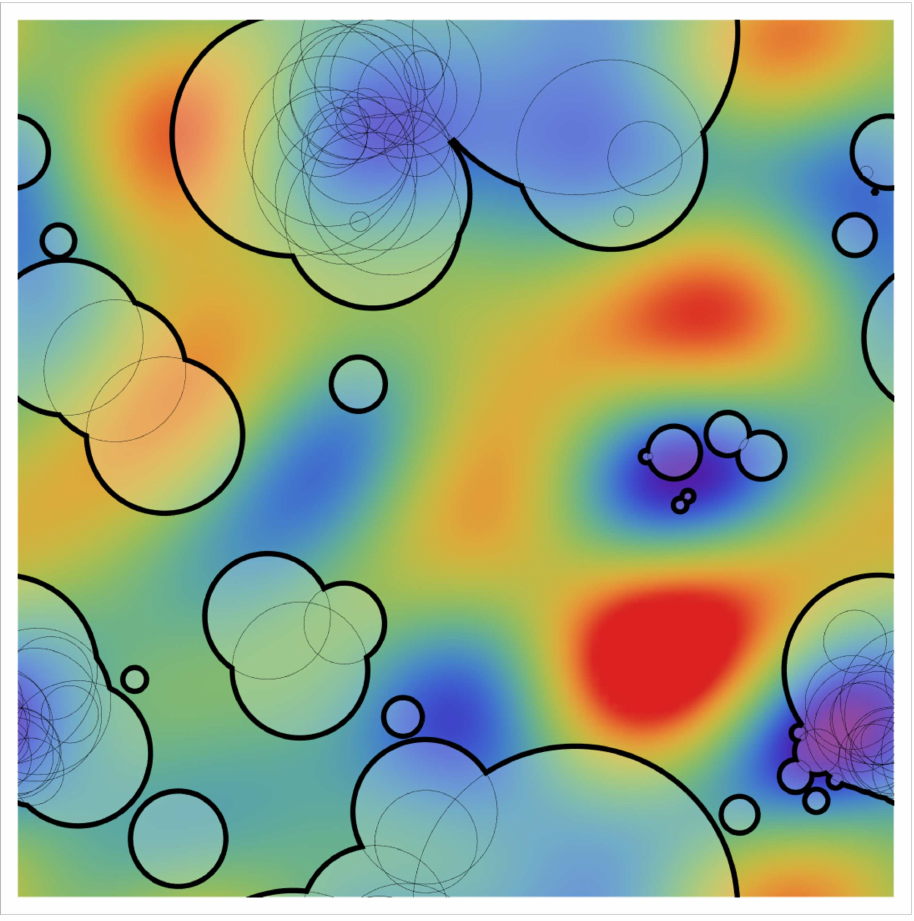}
\includegraphics[width=0.32\textwidth]{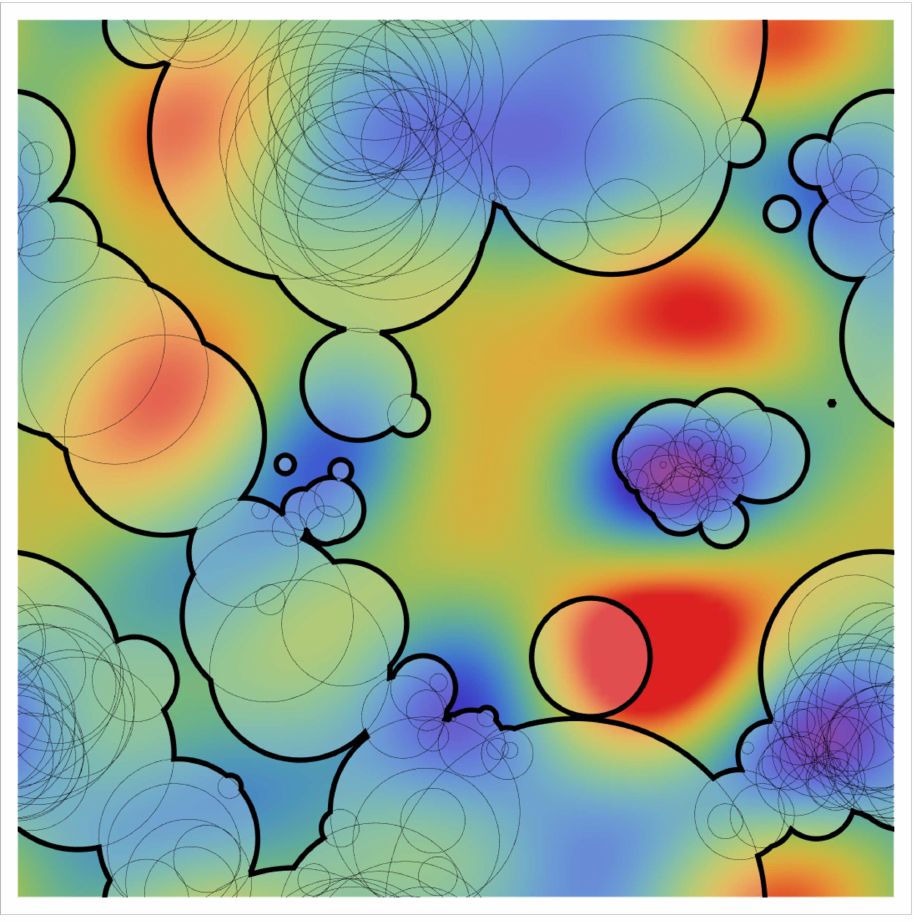}
\includegraphics[width=0.32\textwidth]{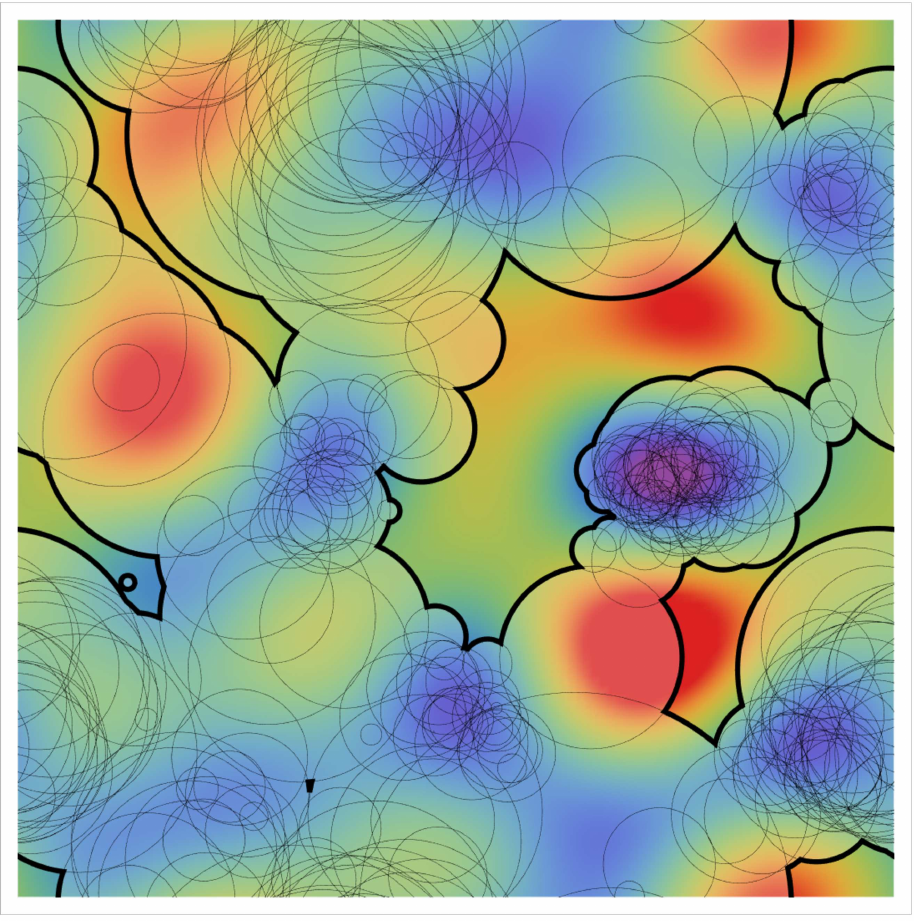}
\vskip 0.1cm
\includegraphics[width=0.32\textwidth]{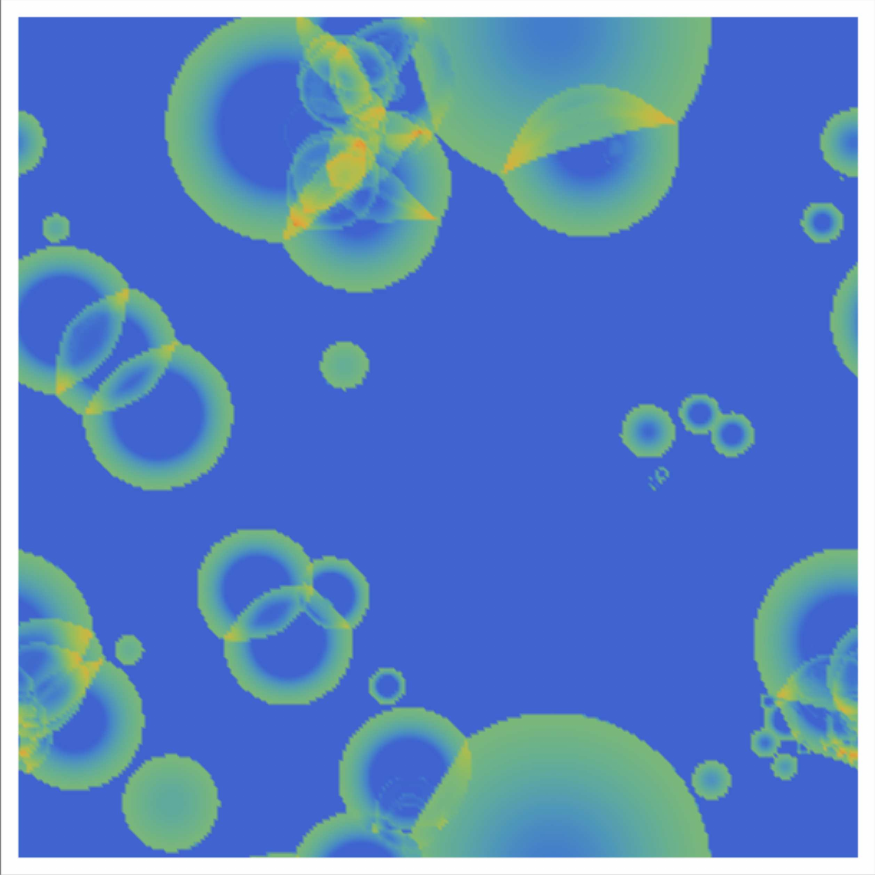}
\includegraphics[width=0.32\textwidth]{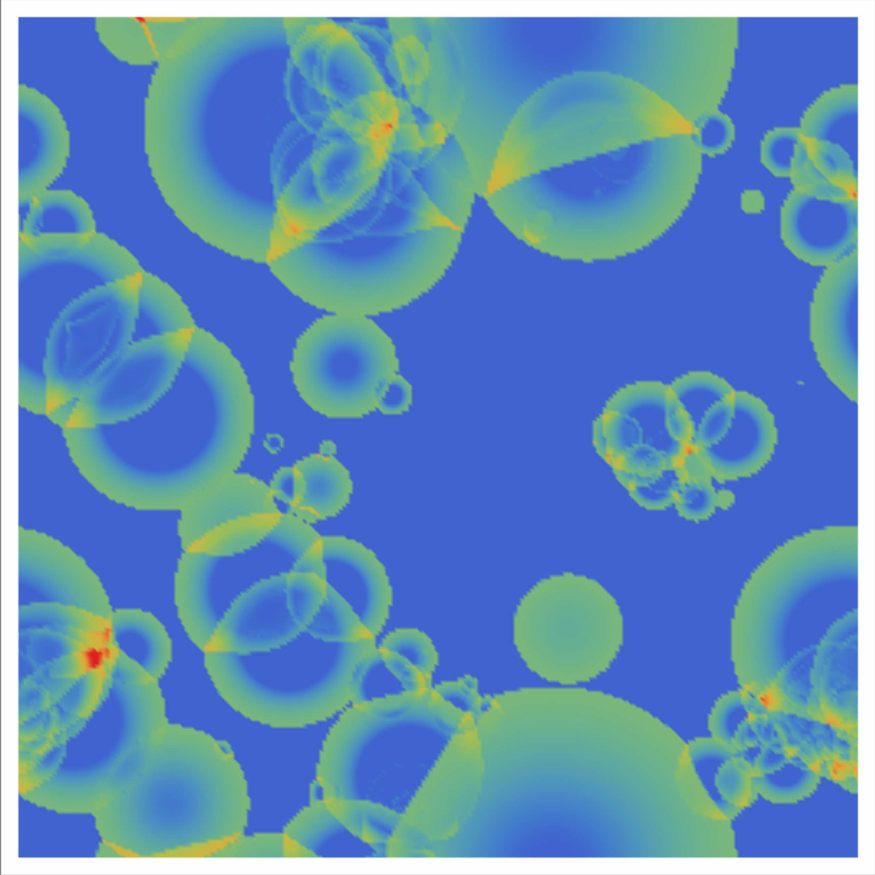}
\includegraphics[width=0.32\textwidth]{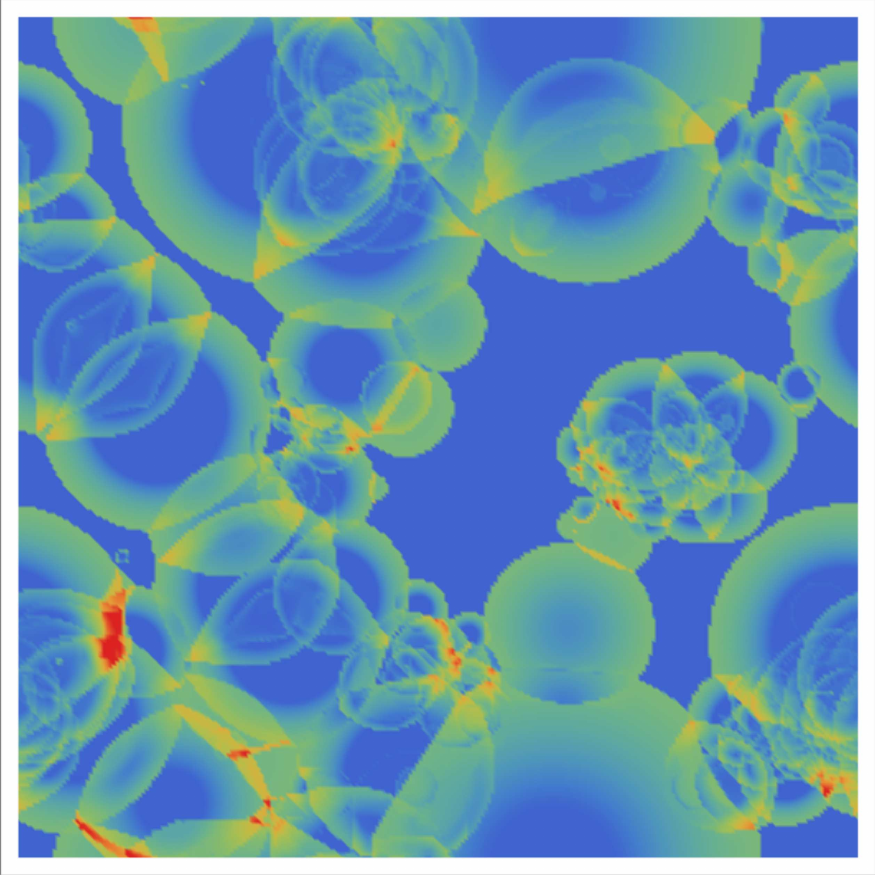}
\caption{
\emph{Top:} Distribution of bubbles in the presence of temperature fluctuations.
We used $L = 40/\beta$, $k_* = 4 \times (2\pi/L)$, and $\sigma = 3$, and the three time slices are $t = - 6 / \beta$, $- 5 / \beta$, and $- 4 / \beta$ from left to right.
The thin circles are causal cones from each nucleation point, while the thick line is the union of the causal cones.
\emph{Bottom:} Fluid velocity distribution calculated from the bubble distribution in the top row.
}
\label{fig:simulation}
\end{figure}
%%%%%%%%%%

%%%%%%%%%%%%%%%%%%%%%%%%%%%%%%%%%%%%%%%%%%%%%%%%%%
\paragraph{Dynamics of the fluctuations.}
%%%%%%%%%%%%%%%%%%%%%%%%%%%%%%%%%%%%%%%%%%%%%%%%%%

In order to calculate the dynamics for $\delta T$, we consider on top of the metric~(\ref{eq:metric}) scalar perturbations in the Newtonian gauge without anisotropic stress. In that case, the single (scalar) degree of freedom is the gravitational potential $\Phi$, for which the equation of motion reads
\begin{align}
	\Phi'' + 3(1+\omega)\mathcal H \Phi'+ \omega k^2 \Phi =0\,,
\end{align}
where the prime denotes a derivative with respect to conformal time $\eta$, $\omega$ is the equation of state parameter and $\mathcal H = a'/a$. This equation holds for adiabatic perturbations and under the assumption that $\omega$ is constant. During radiation domination $\omega = 1/3$ and $\mathcal H = 1/\eta$. For subhorizon modes the radiation density perturbation $\delta_r = \delta \rho_r/\bar \rho_r$ is related to the gravitational potential via \cite{Maggiore:2018sht} 
\begin{align}
	\delta_r \simeq \frac{2}{3} (k \eta)^2 \Phi\,, 
\end{align}
and therefore satisfies
\begin{align}
	\delta''_r - c_s^2 \nabla^2 \delta_r \simeq 0\,,
\end{align}
where we used that $\omega \sim c_s^2$ for a constant equation of state and neglected contributions suppressed by $1/ (k \mathcal H)^2$. For radiation, the temperature perturbation is related to the density perturbation as
\begin{align}
	\frac{\delta T}{\overline{T}} = \frac 1 4 \delta_r\,,
\end{align}
 so $\delta T/\overline{T}$ satisfies 
 \begin{align}
 	\left(\frac{\delta T}{\overline{T}}\right)'' - c_s^2 \nabla^2 \frac{\delta T}{\overline{T}} \simeq 0\,.
 \end{align}
This relation demonstrates that a temperature perturbation oscillates with approximately constant amplitude after it enters the horizon. As the time between the onset and completion of the phase transition is typically much smaller than the Hubble time, we can neglect the difference between conformal and cosmic time. Note that we have neglected the damping and noise terms arising from thermal fluctuations that are considered in \cite{Jackson:2018maa}.
 
We implemented in our algorithm a time-dependent temperature grid and later in the text we see how the dynamics of the density waves may affect the number of bubbles nucleated. In particular, the impact of the fluctuations will be quite different in the limit of very small (IR) or large (UV) wavenumbers of the fluctuations.
 
%%%%%%%%%%
\begin{figure}[t]
\centering
\includegraphics[width=0.3\textwidth]{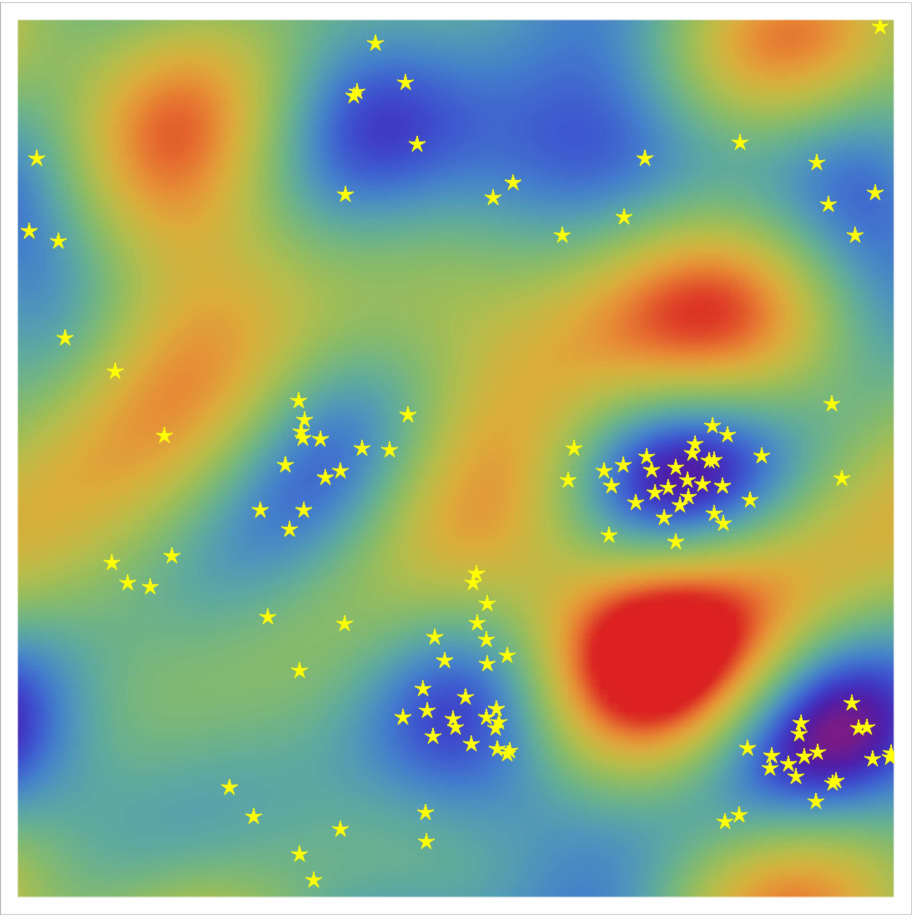}
\hskip 1cm
\includegraphics[width=0.45\textwidth]{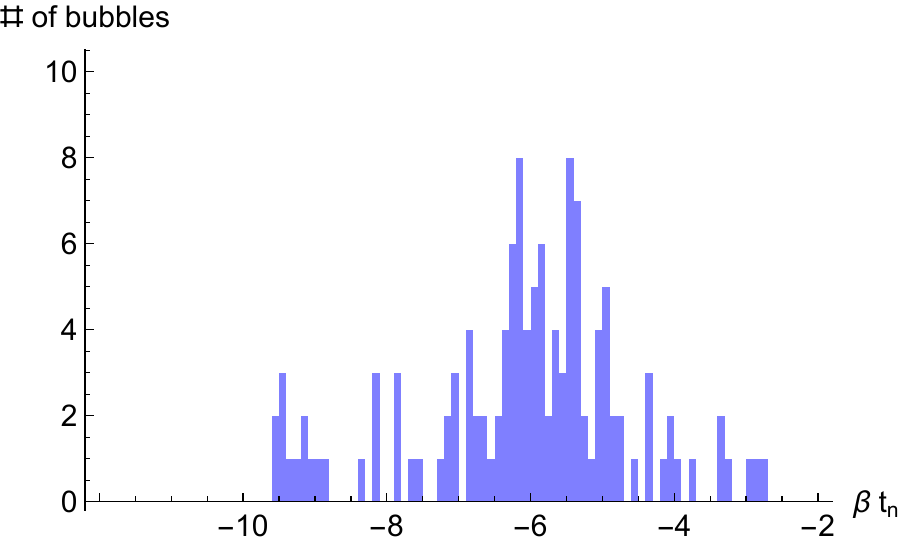}
\vskip 0.5cm
\includegraphics[width=0.3\textwidth]{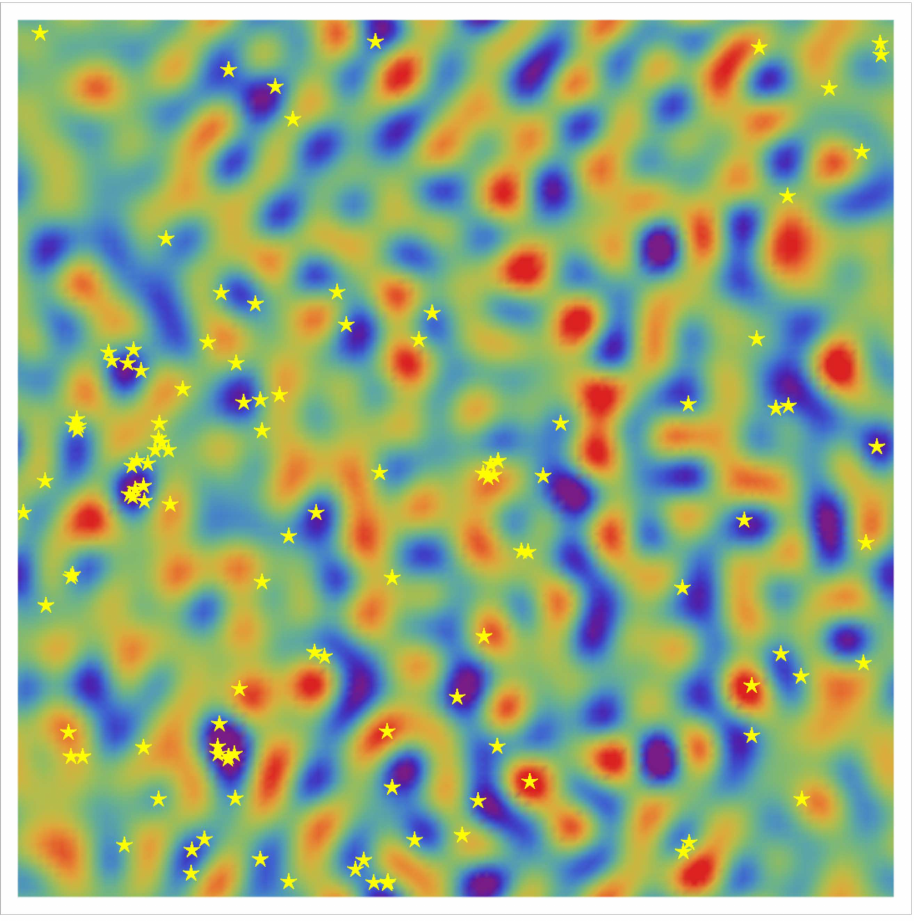}
\hskip 1cm
\includegraphics[width=0.45\textwidth]{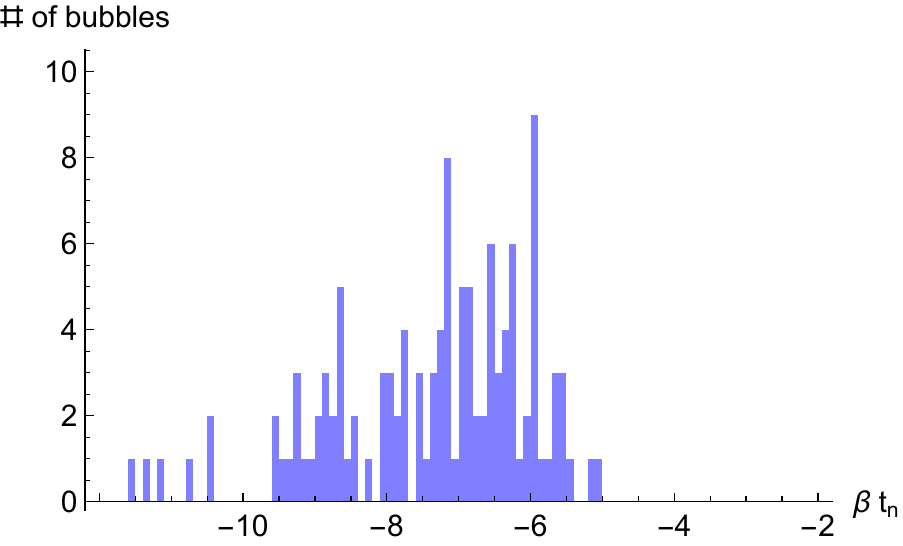}
\caption{
Position and time distribution of the nucleation points in the presence of temperature fluctuations for different $k_*$.
The typical wavenumber is $k_* = 4 \times (2 \pi / L)$ and $k_* = 16 \times (2 \pi / L)$ for the top and bottom row, respectively, while $L = 40 / \beta$ and $\sigma = 3$ for both.
The left panels display the nucleation points for the whole simulation time around the $z$ slice plotted (within $\Delta z = \pm 1 / \beta$), while the right panel shows the nucleation time distribution around these $z$ slices.
In the left panel, the density fluctuations are plotted in color at the typical bubble nucleation time ($t \simeq - 6 / \beta$ and $t \simeq - 7 / \beta$ for the top and bottom panels, respectively). 
}
\label{fig:density}
\end{figure}
%%%%%%%%%%

%%%%%%%%%%%%%%%%%%%%%%%%%%%%%%%%%%%%%%%%%%%%%%%%%%
\paragraph{IR vs. UV fluctuations.}
%%%%%%%%%%%%%%%%%%%%%%%%%%%%%%%%%%%%%%%%%%%%%%%%%%

We compare the effect of large-scale and small-scale fluctuations on bubble nucleation. Since, without fluctuations, the average size of the bubbles (that is related to $\beta$) is the only relevant length scale in the problem, we call UV temperature fluctuation those which have a correlation scale $k_* \gg \beta$ and the IR fluctuations those for which $k_* < \beta$.

We will see that for IR fluctuations, nucleation will mainly take place in the cold spots, which leads to enhanced GW signals.
In Fig.~\ref{fig:simulation} we show how the distribution of bubbles is biased by the temperature fluctuations.
In these plots we used $L = 40 / \beta$, $k_* = 4 \times (2 \pi / L)$, and $\sigma = 3$.
We see that the bubbles start to nucleate at the cold spots first, and these bubbles expand up to the typical size of the temperature fluctuation.
Therefore we expect that the ``effective bubble size" is determined by the typical size of the hot and cold spots, as long as the shift in nucleation time induced by the temperature fluctuation is comparable or larger than the typical timescale for the completion of the transition, $\sigma \gtrsim H_* / \beta$. Notice that we use dynamical temperature 
fluctuations with $c_s^2 = 1/3$. In the deep-IR limit $k_\ast < H_\ast$ (or $k_\ast < L^{-1}$ for our simulation), we of course expect the system to be oblivious to temperature fluctuations and reproduce the case $\delta T = 0$.

\begin{figure}
\centering
\includegraphics[width=0.45\textwidth]{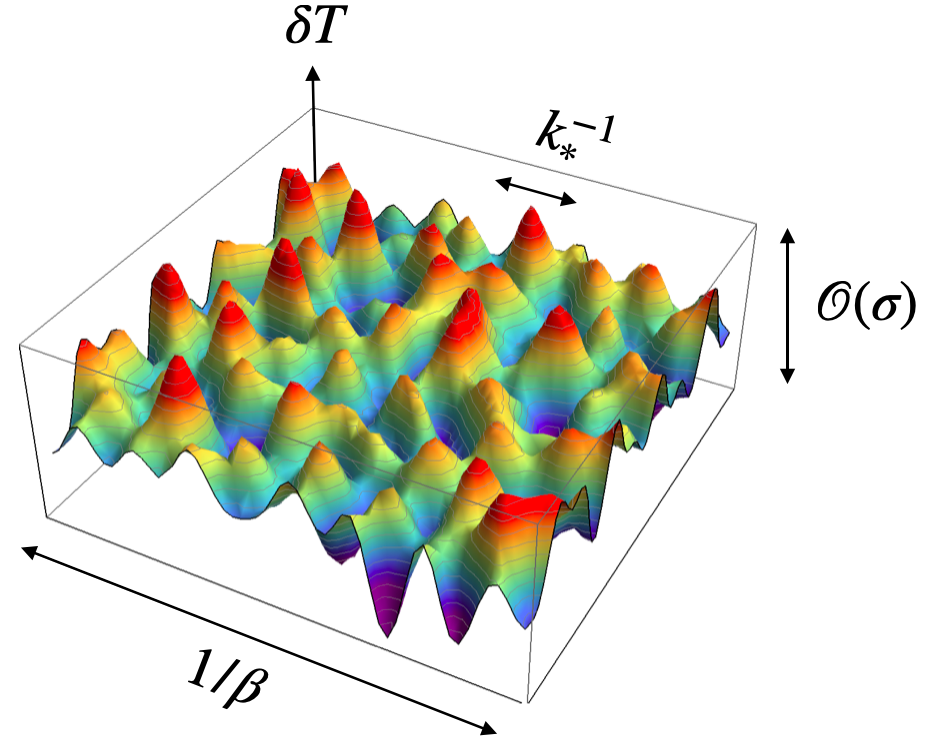} 
\caption{
Illustration for why UV fluctuations do not give significant change in the bubble distribution. The typical change in nucleation position is of the order of $k_*^{-1}$, which is much smaller than the typical bubble size, which scales as $\propto 1/\beta$.
}
\label{fig:UV}
\end{figure}

For UV fluctuations, meaning $k_* \gg \beta$ and $\sigma$ fixed, the spatial distribution of nucleated bubbles is hardly affected. 
To illustrate this point, in Fig.~\ref{fig:density} we compare how the location of bubble nucleation is biased by the temperature fluctuations for large-scale and small-scale density fluctuations.
We indeed see that the impact is smaller for UV fluctuations.
This can be understood from the fact that in the limit of $k_* \rightarrow \infty$ any finite volume element has infinitely many hot and cold spots. Note that the nucleation history experiences a collective time shift
of the bubble nucleations, given by
\be
\Delta t = 
\frac{\sigma^2}{2 \beta}\,.
\label{eq:tshift}
\ee
This relation follows from the fact that $\delta \tilde{T}$ is (approximately) Gaussian which leads to the relation
\be
\frac1{V} \int d^3x \, \exp(\delta \tilde{T}) = \exp(\sigma^2/2) \, .
\ee
In this limit there is no net effect on the bubble distribution.
In fact, while large-scale temperature fluctuations induce a bias in the nucleation positions (see Fig.~\ref{fig:density}, top left), small-scale fluctuations leave essentially no effect in the spatial distribution of the nucleation points (see Fig.~\ref{fig:density}, bottom left panel and Fig.~\ref{fig:UV}).

We can then conclude that the system will only be sensitive to temperature fluctuations with $H_* < k_* < \beta$. Some more analytical estimates of the impact of UV and IR modes on the nucleation 
history are given in Appendix~\ref{app:analytic}. 

%%%%%%%%%%
%%%%%%%%%%

%%%%%%%%%%%%%%%%%%%%%%%%%%%%%%%%%%%%%%%%%%%%%%%%%%
\paragraph{Comments.}
%%%%%%%%%%%%%%%%%%%%%%%%%%%%%%%%%%%%%%%%%%%%%%%%%%

In passing we comment on several points.
First, the GW production mechanism we discuss in this paper is different from the GW production from the density perturbations themselves. The latter was studied for example in \cite{Ananda:2006af,Baumann:2007zm}. In this case, the gravitational potential is directly related to the density perturbation and as a result, the gravitational wave power spectrum scales directly with the power spectrum of the primordial density perturbation
\begin{align}
\Omega_{\rm GW}
&\propto
{\cal P}_\zeta^2\,,
\label{eq:gwperts}
\end{align}
with ${\cal P}_\zeta$ being the dimensionless power spectrum of the curvature perturbation $\zeta$.
On the other hand, the GW signal from sound waves behaves as
\begin{align}
\Omega_{\rm GW, sw}
&\propto
\left( \frac{\kappa \alpha}{1 + \alpha} \right)^2
R_* H_* \frac{1}{v_w}\,.
\end{align}
 The effect of the temperature fluctuations on the phase transition is to enhance the typical value of $R_*$ and thereby the amplitude of the gravitational wave signal. The strength of the effect increases with $\sigma$, and thus with ${\cal P}_\zeta^2$, but the parametric dependence is different from Eq.~(\ref{eq:gwperts}). We will see this in the numerical results of the 
next section.

Second, the limit where one takes either $\sigma$ or $k_*$ to infinity is challenging to simulate. When $\sigma$ becomes large, the nucleation probability in space becomes sharply peaked in isolated locations since the probability depends exponentially on $\sigma$. The nucleation time of the very first bubble depends hence on the tail of the Gaussian distribution.
The correct sampling with large $\sigma$ would therefore demand to exponentially increase the grid resolution in order to track the bubble nucleation distribution correctly. At the same time, when $k_*$ becomes too large, resolving $k_*$ on the simulation grid in combination with many bubbles on the lattice gets prohibitively expensive.

%%%%%%%%%%%%%%%%%%%%%%%%%%%%%%%%%%%%%%%%%%%%%%%%%%
\section{Results}
\label{sec:Results}
%%%%%%%%%%%%%%%%%%%%%%%%%%%%%%%%%%%%%%%%%%%%%%%%%%

In this section we present results of numerical simulations to study the effects laid out in the previous section.
We discuss the case in which the temperature fluctuations are dynamical ($c_s^2 = 1/3$) but also the static case ($c_s^2 = 0$),
since both lead to quite different results in bubble nucleation histories.
The static limit provides a simplified scenario that provides physical intuition and is easier to simulate. 
In the following simulations we set $\Gamma_* = \beta^4$ and $t_* = 0$ without loss of generality.

\paragraph{Bubble statistics.}

\begin{figure}
\centering
\includegraphics[width=0.45\textwidth]{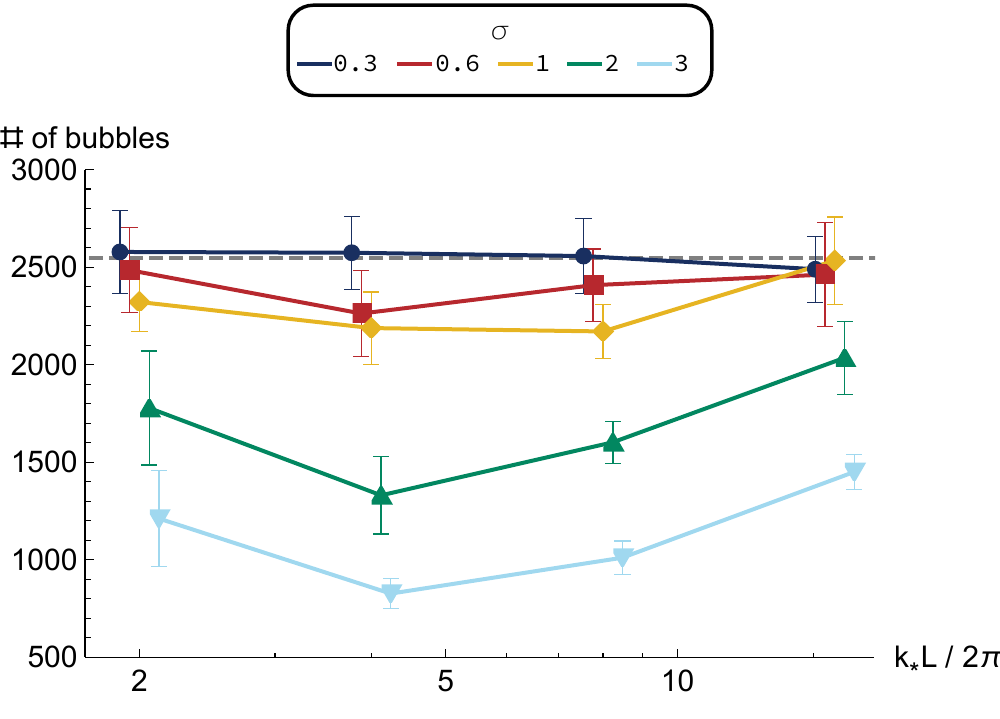} 
\includegraphics[width=0.45\textwidth]{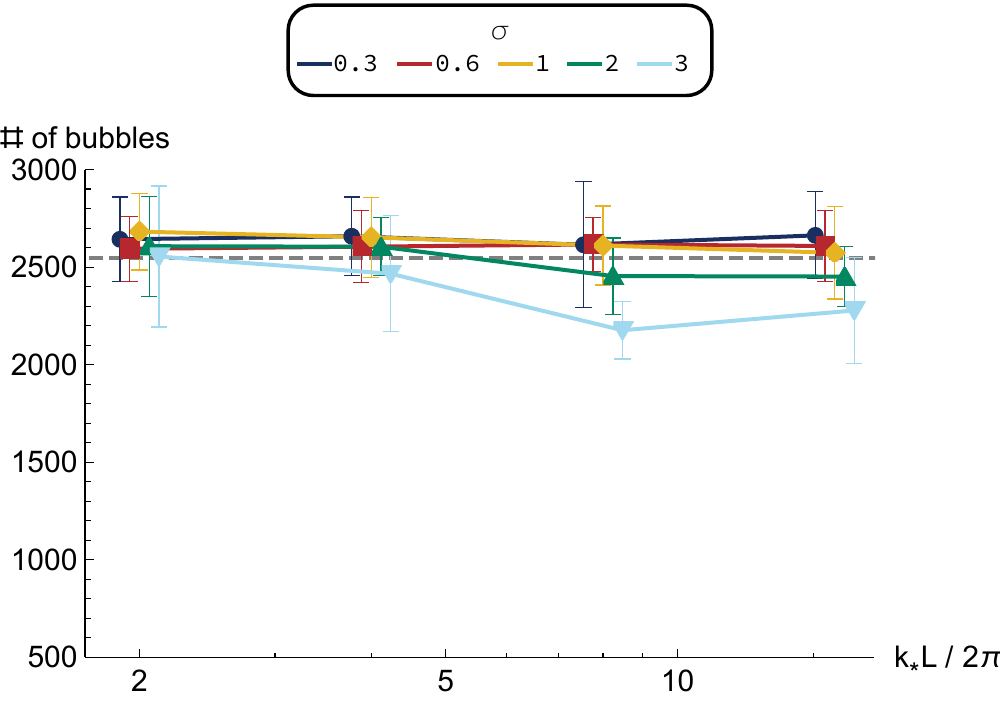} 
\caption{
Number of bubbles for different amplitudes and wavenumbers of the density fluctuation.
The box size is $L = 40 / \beta$.
The dashed line is the prediction without the fluctuations $N = (\beta L)^3 / 8 \pi$.
\emph{Left:} Time-independent fluctuation $c_s^2 = 0$.
\emph{Right:} Time-dependent fluctuation $c_s^2 =1/3$. The error bars are calculated as the variance of 10 simulations.
}
\label{fig:Nbubble}
\end{figure}

We first study the bubble statistics and nucleation history before discussing GW production.
In Fig.~\ref{fig:Nbubble} we plot the total number of bubbles that is formed until the phase transition completes, $N_b$, for different values of $k_*$ and $\sigma$ averaged over $10$ realizations in a box of $L = 40 / \beta$.
The left and right panel is for $c_s^2 = 0$ and $c_s^2 = 1/3$, respectively.
While for both cases the small $\sigma$ limit reproduces the theoretical prediction without fluctuation $N_b = (\beta L)^3 / 8 \pi$ shown in the gray-dashed lines, the behavior for larger $\sigma$ is quite different.

For $c_s^2 = 0$, the relative temperature fluctuation remains constant at any location.
Because of this, for large $\sigma$, bubbles do not nucleate in the hot spots before the bubbles from the cold spots arrive.
Since the typical nucleation time difference between hot and cold spots is given by $\beta \Delta t_n \sim \sigma^2/2$, this effect is stronger for larger $\sigma$.
We clearly see this tendency in the left panel of Fig.~\ref{fig:Nbubble}.
We also observe that the number of bubbles tends to increase for larger $k_*$.
As explained in Sec.~\ref{sec:Idea}, the effect of temperature fluctuation vanishes in the $k_* \to \infty$ limit for a fixed $\sigma$. This argument holds true irrespectively of the time dependence of the fluctuation and we observe it in our simulations.

For $c_s^2 = 1/3$, we do not observe any significant decrease in the number of bubbles. However, just like the case with $c_s^2=0$, fewer bubbles nucleate in the hot spots. In the cold spots the movement of the temperature troughs leads to the nucleation of many smaller bubbles that merge to form a larger effective bubble, see Fig.~\ref{fig:simulation}. Hence, the number of bubbles is almost unaffected, compared to the case without fluctuations, but the spatial distribution \emph{is}. As we will see below, significant GW enhancement occurs even in such cases.
In other words, the bubble count is not a good indicator of the GW signal. 

%%%%%%%%%%
%%%%%%%%%%

\paragraph{Nucleation time distribution.}

%%%%%%%%%%
\begin{figure}
\centering
\includegraphics[width=0.32\textwidth]{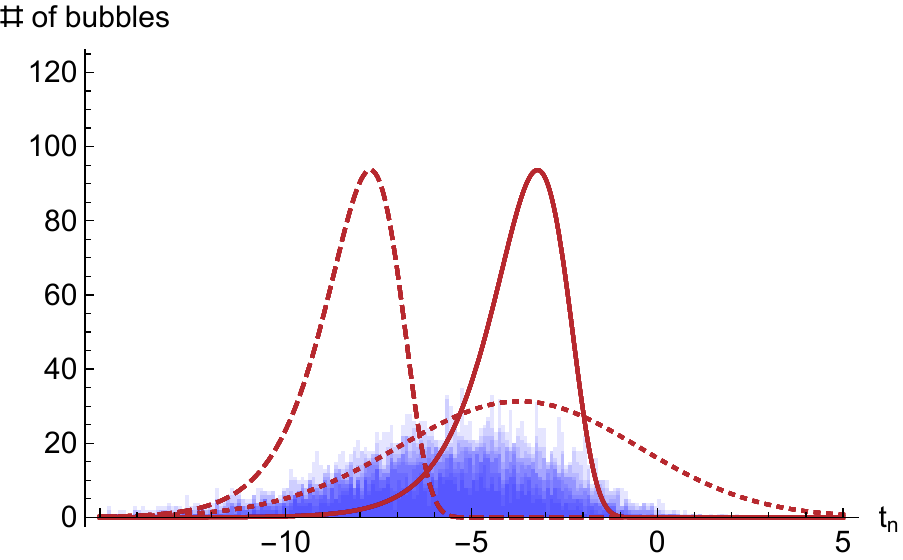}
\includegraphics[width=0.32\textwidth]{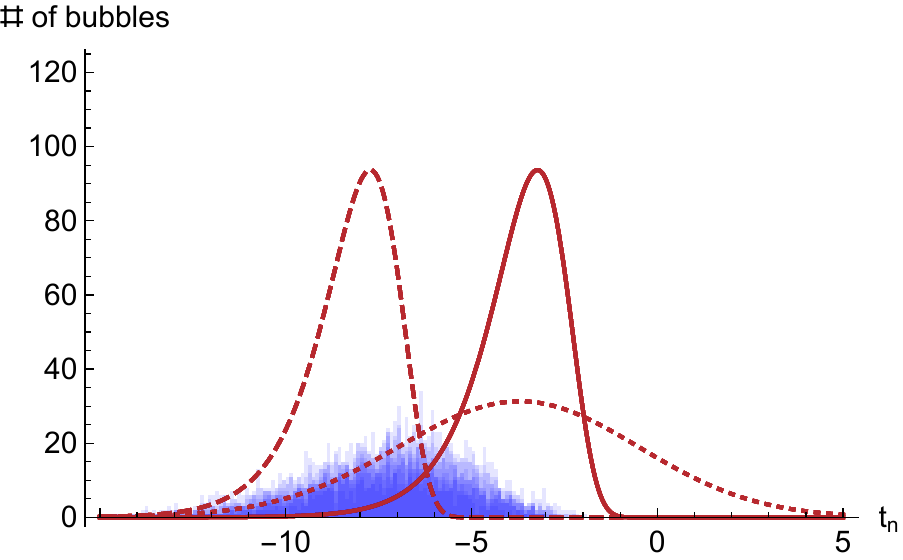}
\includegraphics[width=0.32\textwidth]{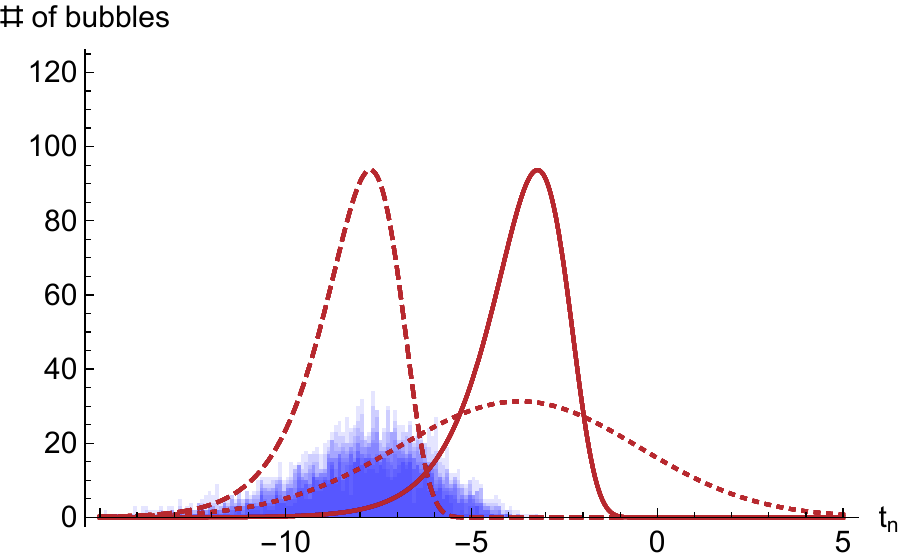}
\caption{
Nucleation time distribution with different $\sigma$ and $k_*$ for $c_s = 0$.
In this plot we take $\sigma = 3$, and $k_* = (2, 4, 6) \times (2 \pi / L)$ from left to right.
The box size is $L = 40 / \beta$, and we overlay $10$ nucleation histories for each panel.
The red lines are $P_{{\rm nuc}, \delta \tilde{T} = 0}$ (red-solid), $P_{\rm nuc, UV}$ (red-dashed), and $P_{\rm nuc, IR}$ (red-dotted) when the bin size is $\Delta t_n = 0.1 / \beta$.
The main feature to observe is that the number of bubbles that are nucleated at late times is significantly reduced. 
}
\label{fig:tNuc_indep}
\vskip 1cm
\centering
\includegraphics[width=0.32\textwidth]{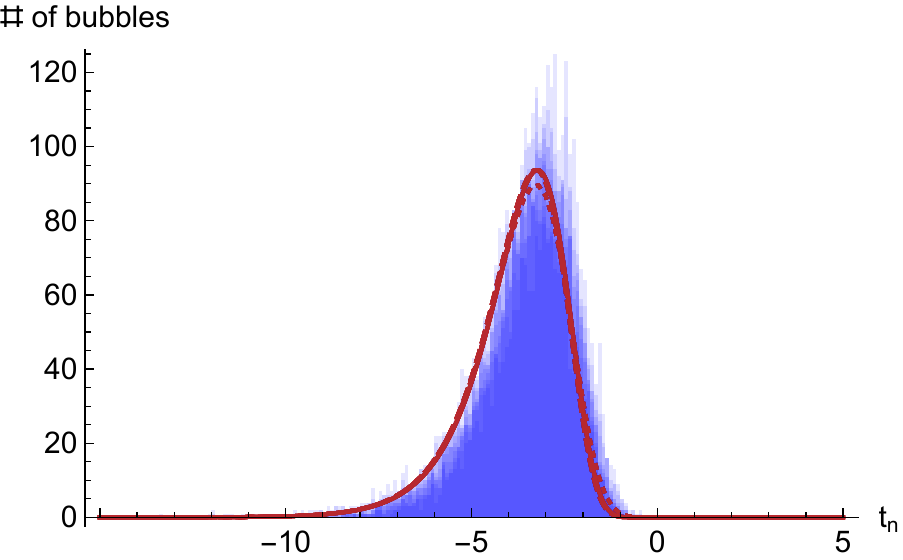}
\includegraphics[width=0.32\textwidth]{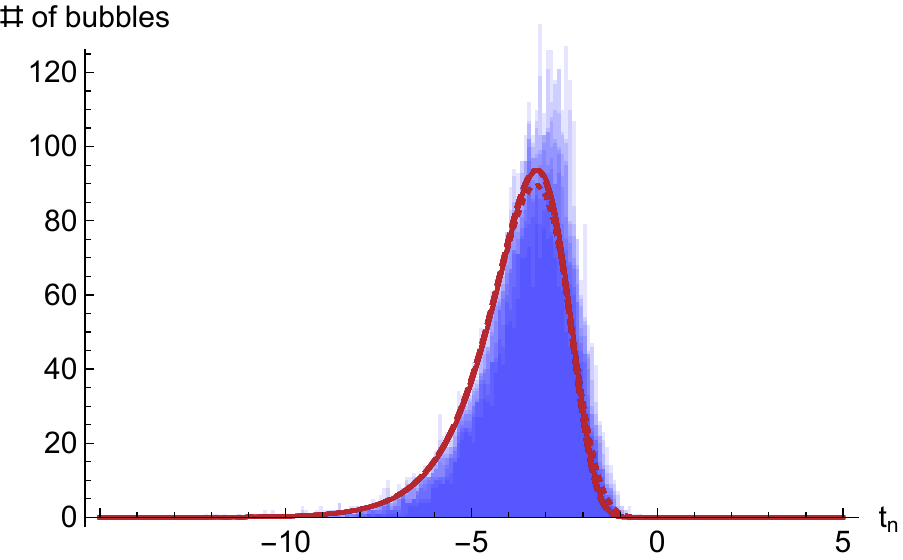}
\includegraphics[width=0.32\textwidth]{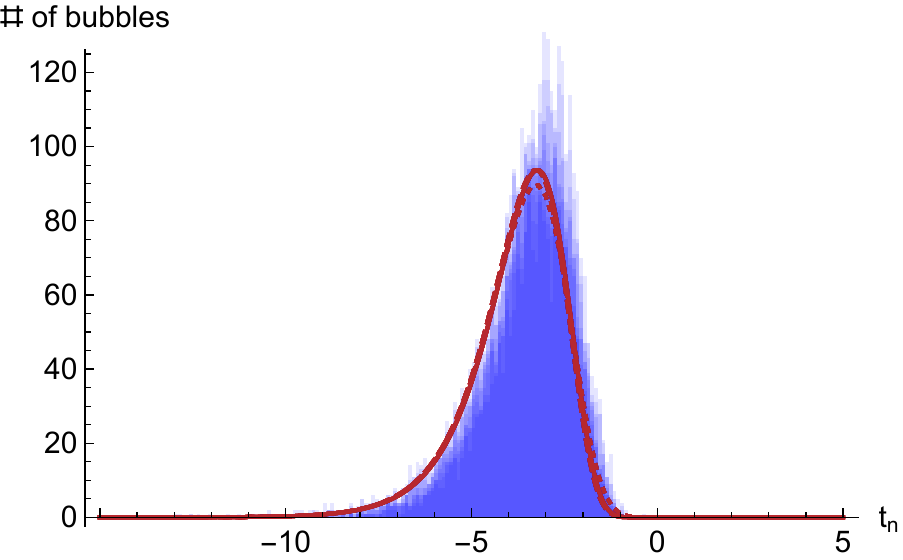}
\includegraphics[width=0.32\textwidth]{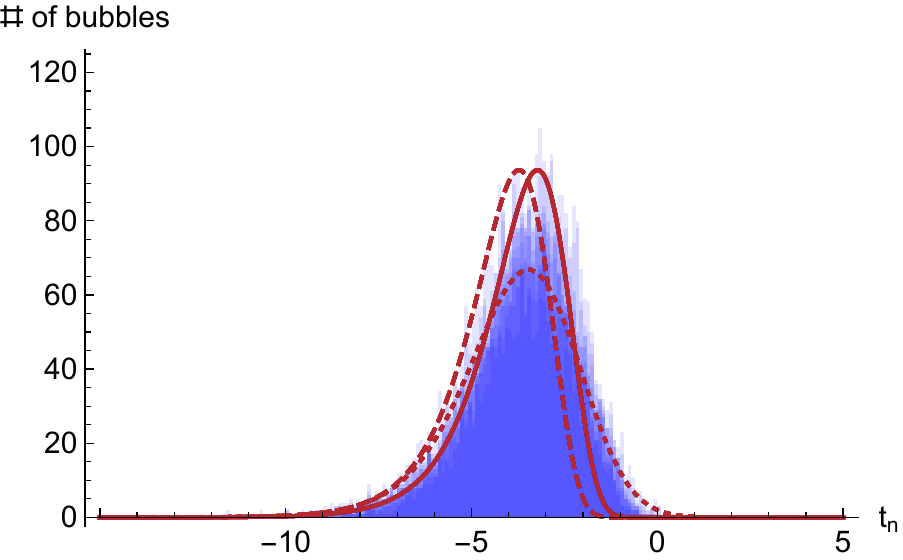}
\includegraphics[width=0.32\textwidth]{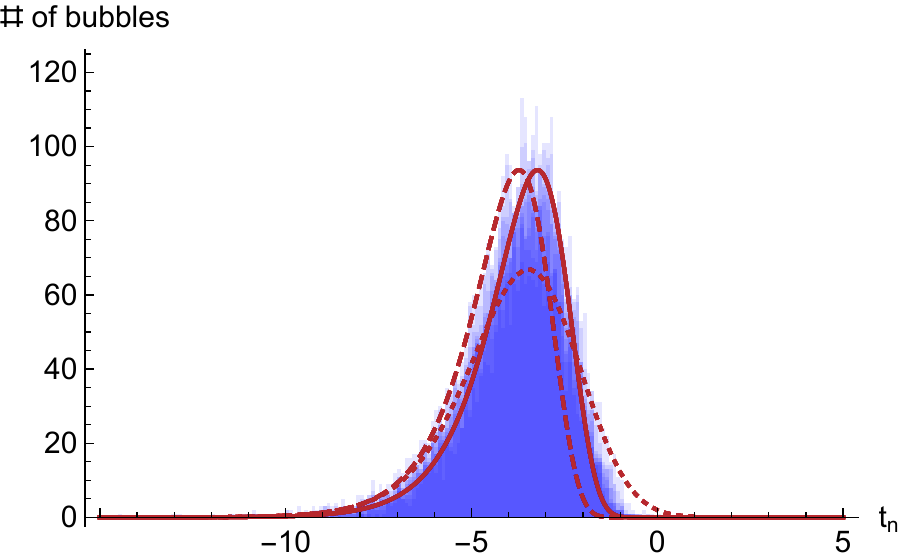}
\includegraphics[width=0.32\textwidth]{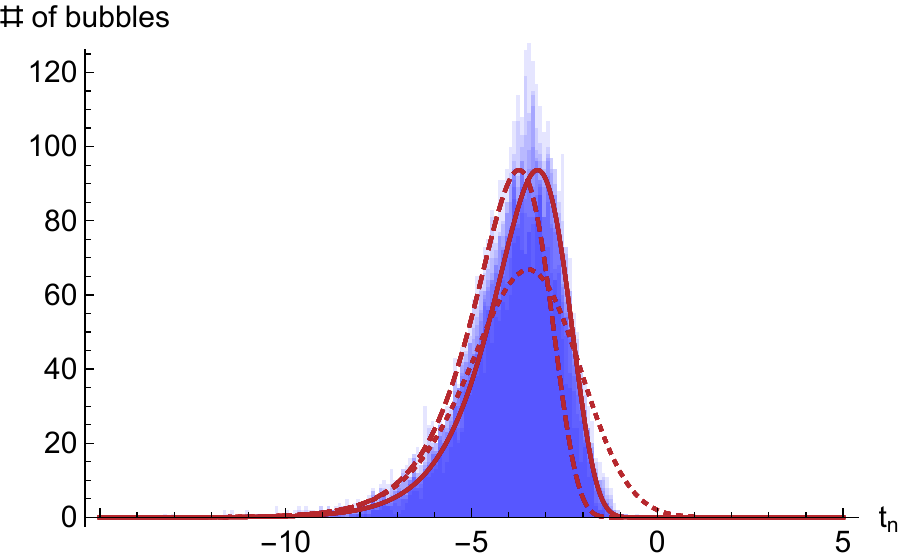}
\includegraphics[width=0.32\textwidth]{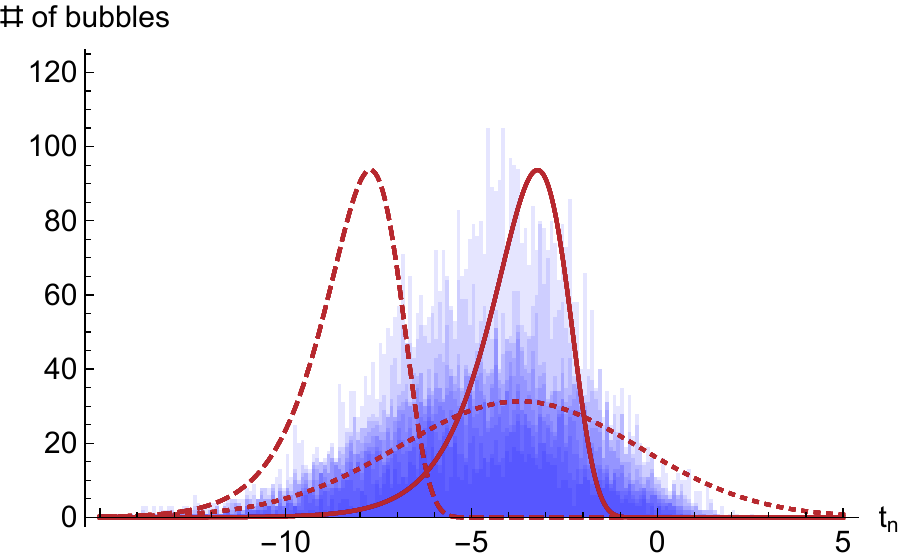}
\includegraphics[width=0.32\textwidth]{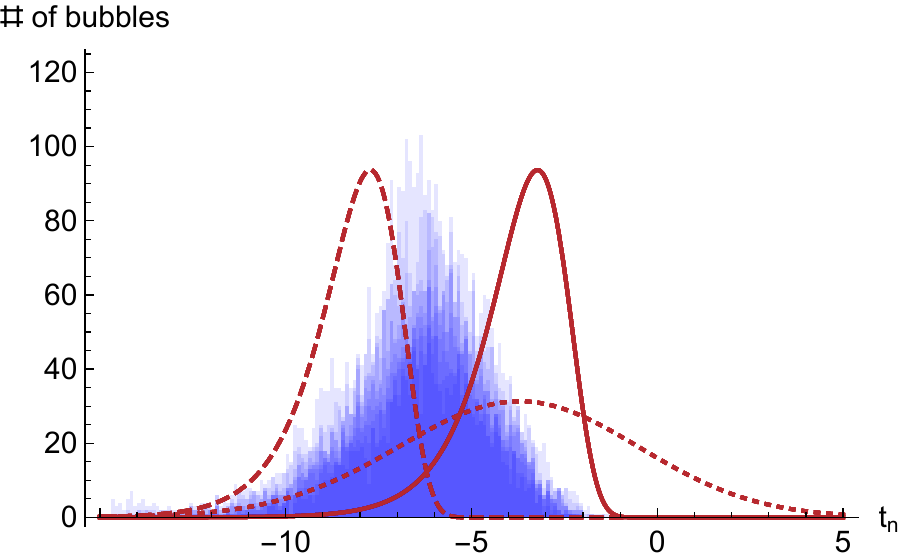}
\includegraphics[width=0.32\textwidth]{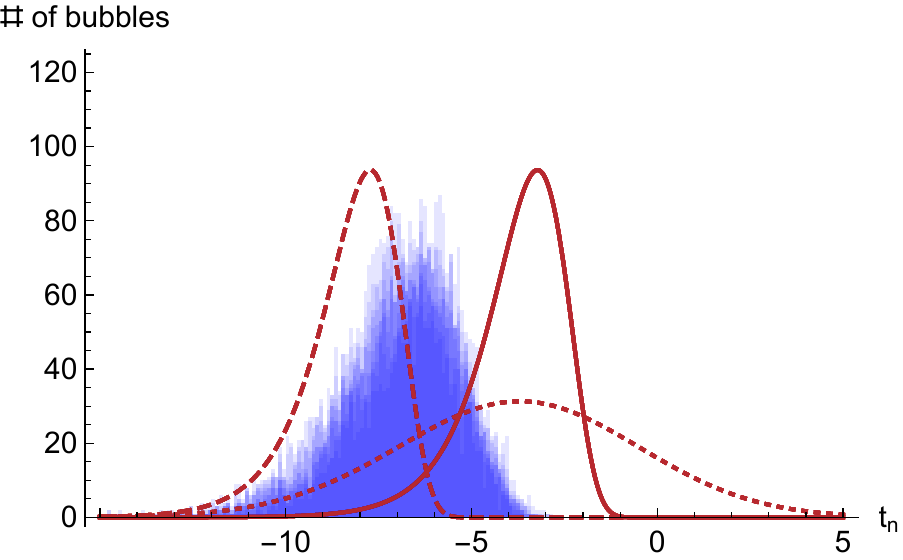}
\caption{
Nucleation time distribution with different $\sigma$ and $k_*$ for $c_s^2 =1/3$.
In this plot we take $\sigma = 0.3$, $1$, and $3$ from top to bottom and $k_* = (2, 4, 6) \times (2 \pi / L)$ from left to right, while other parameters and the red lines are the same as Fig.~\ref{fig:tNuc_indep}.
The bottom row displays exactly the same parameter choices as Fig.~\ref{fig:tNuc_indep} (except for $c_s$)}.

\label{fig:tNuc_dep}
\end{figure}
%%%%%%%%%%

We now compare the time distribution for the bubbles nucleated in the presence of temperature fluctuations and compare it to the $\delta T = 0$ scenario.
Without the temperature fluctuation, the nucleation time distribution is given by
\begin{align}
P_{n, \delta \tilde{T} = 0} (t_n)
&= 
8 \pi \Gamma_* e^{\beta (t_n - t_*) - 8 \pi \Gamma_* e^{\beta (t_n - t_*)}}\,,
\end{align}
while in the IR and UV limits
\begin{align}
P_{n, {\rm IR}} (t_n)
&= 
\frac{1}{\sqrt{2 \pi \sigma^2}}
\int d (\beta \delta t_n)~
e^{- \frac{(\beta \delta t_n)^2}{2 \sigma^2}}
P_{n, \delta \tilde{T} = 0} (t_n + \delta t_n)\,,
\nn \\[0.2cm]
P_{n, {\rm UV}} (t_n)
&= 
P_{n, \delta \tilde{T} = 0} \left( t_n + \frac{\sigma^2}{2 \beta} \right)\,.
\label{eq:Plimits}
\end{align}
While we put the derivation in Appendix~\ref{app:analytic}, the interpretation of these expressions is quite straightforward.
In the IR limit, each horizon patch is covered with a constant fluctuation, and thus the net effect is just a time shift for each patch obeying a Gaussian distribution.
In the UV limit, the fluctuations lead to a collective time shift as explained around Eq.~(\ref{eq:tshift}).
These IR and UV limits, as well as the distribution without the fluctuation, are plotted as red lines in Figs.~\ref{fig:tNuc_indep} and \ref{fig:tNuc_dep}.
The red-solid line is the distribution without the fluctuation, while the red-dotted and red-dashed lines are the IR and UV limits, respectively.

The nucleation time distribution for $c_s = 0$ is shown in Fig.~\ref{fig:tNuc_indep}.
In this plot we take $\sigma = 3$, and $k_* = (2, 4, 6) \times (2 \pi / L)$ from left to right.
The box size is $L = 40 / \beta$, and we overlay $10$ nucleation histories for each panel.
The number of bubbles is significantly reduced compared to the case with dynamical fluctuations.
The reason is that since the cold and hot spots are not moving, most regions with a high nucleation probability already
complete the phase transition before additional bubbles have a chance to be nucleated. 

The nucleation time distributions for $c_s^2 = 1/3$ and with different $\sigma$ and $k_*$ are shown in Fig.~\ref{fig:tNuc_dep}.
In this plot we take $\sigma = 0.3$, $1$, and $3$ from top to bottom, and $k_* = (2, 4, 6) \times (2 \pi / L)$ from left to right.
Other parameters and the red lines are the same as Fig.~\ref{fig:tNuc_indep}.
The plots show that the distributions (slowly) approach the limiting distributions as predicted in Eq.~(\ref{eq:Plimits}). As predicted, as the amplitude of the fluctuations grows, the time distribution gets broadened for IR fluctuations. Yet, understanding the effect on the GW spectrum does not only require knowledge of the nucleation history, but also of the bubble distribution in space. We now move to understand how bubble clustering affects the GW spectrum.

%%%%%%%%%%%%%%%%%%%%%%%%%%%%%%%%%%%%%%%%%%%%%%%%%%
\paragraph{Gravitational waves}
\label{subsec:GWs}
%%%%%%%%%%%%%%%%%%%%%%%%%%%%%%%%%%%%%%%%%%%%%%%%%%

%%%%%%%%%%
\begin{figure}
\centering
\includegraphics[width=0.3\textwidth]{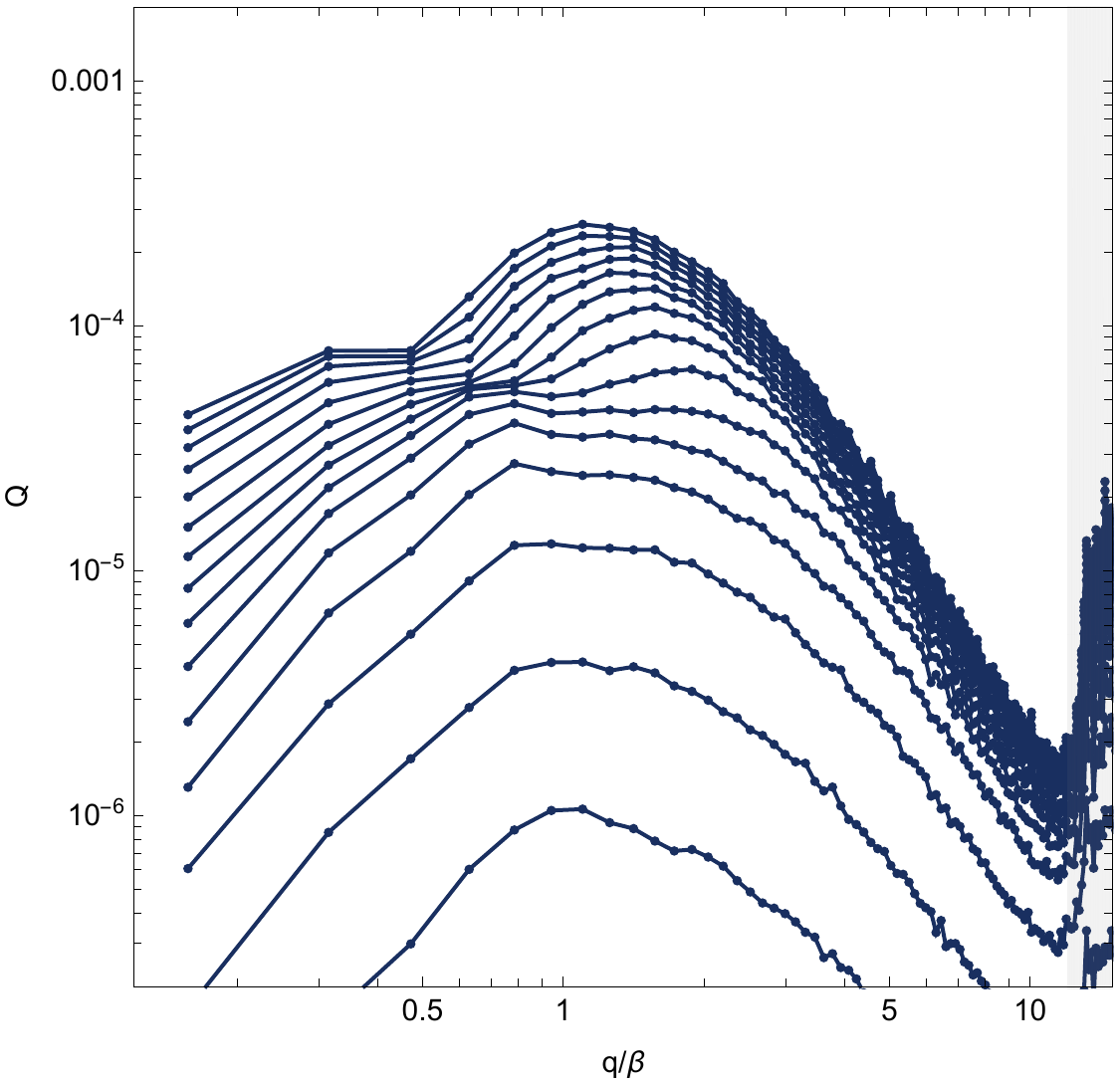}
\caption{
Gravitational-wave spectrum without temperature fluctuations.
We used the box size $L = 40 / \beta$ and wall velocity $v_w = 1$.
Different lines correspond to different time slices up to $t = 10 / \beta$ with time interval $\Delta t = 1 / \beta$.
}
\label{fig:OmegaGW_0}
\vskip 1cm
\includegraphics[width=0.3\textwidth]{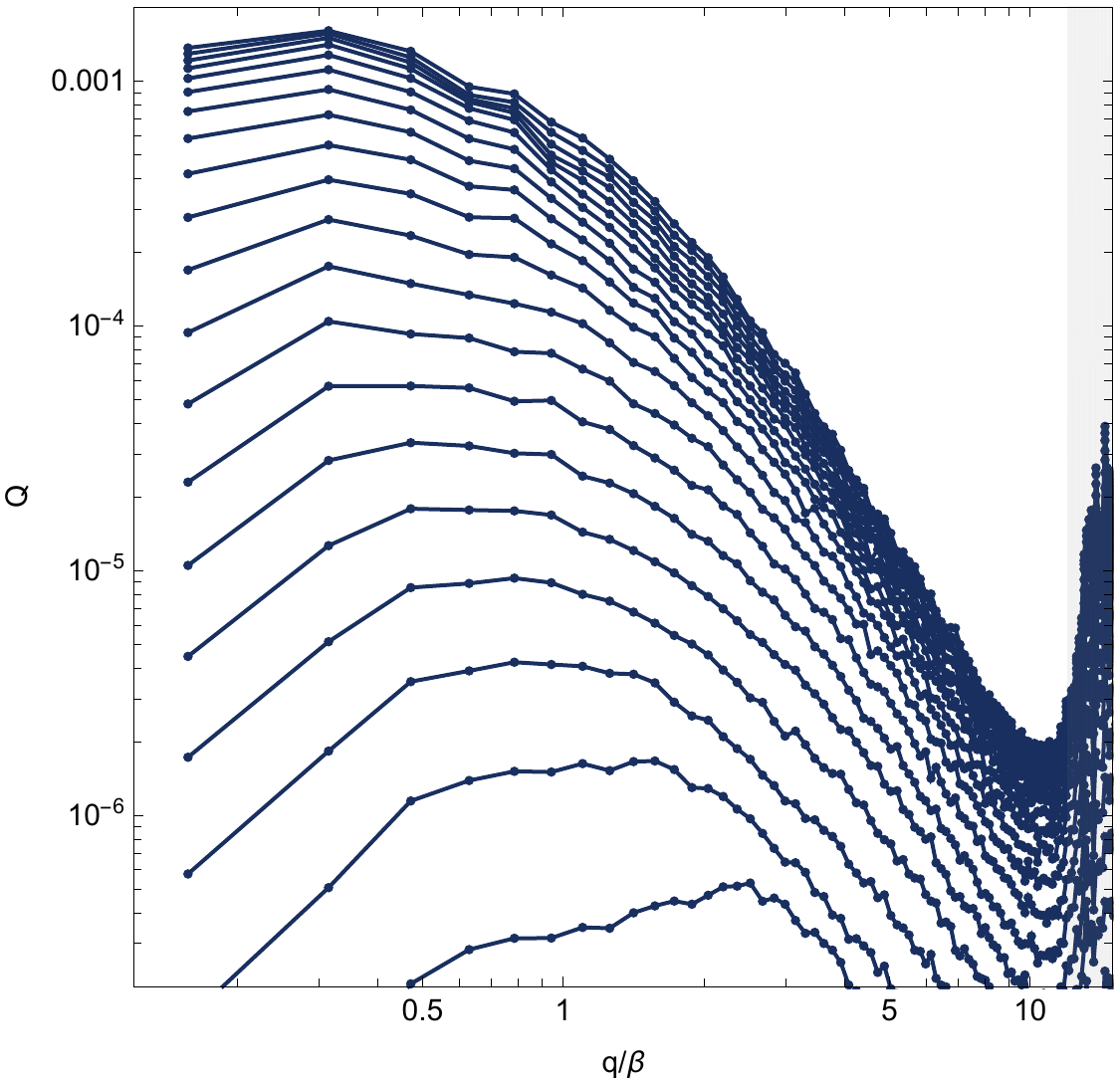}
\includegraphics[width=0.3\textwidth]{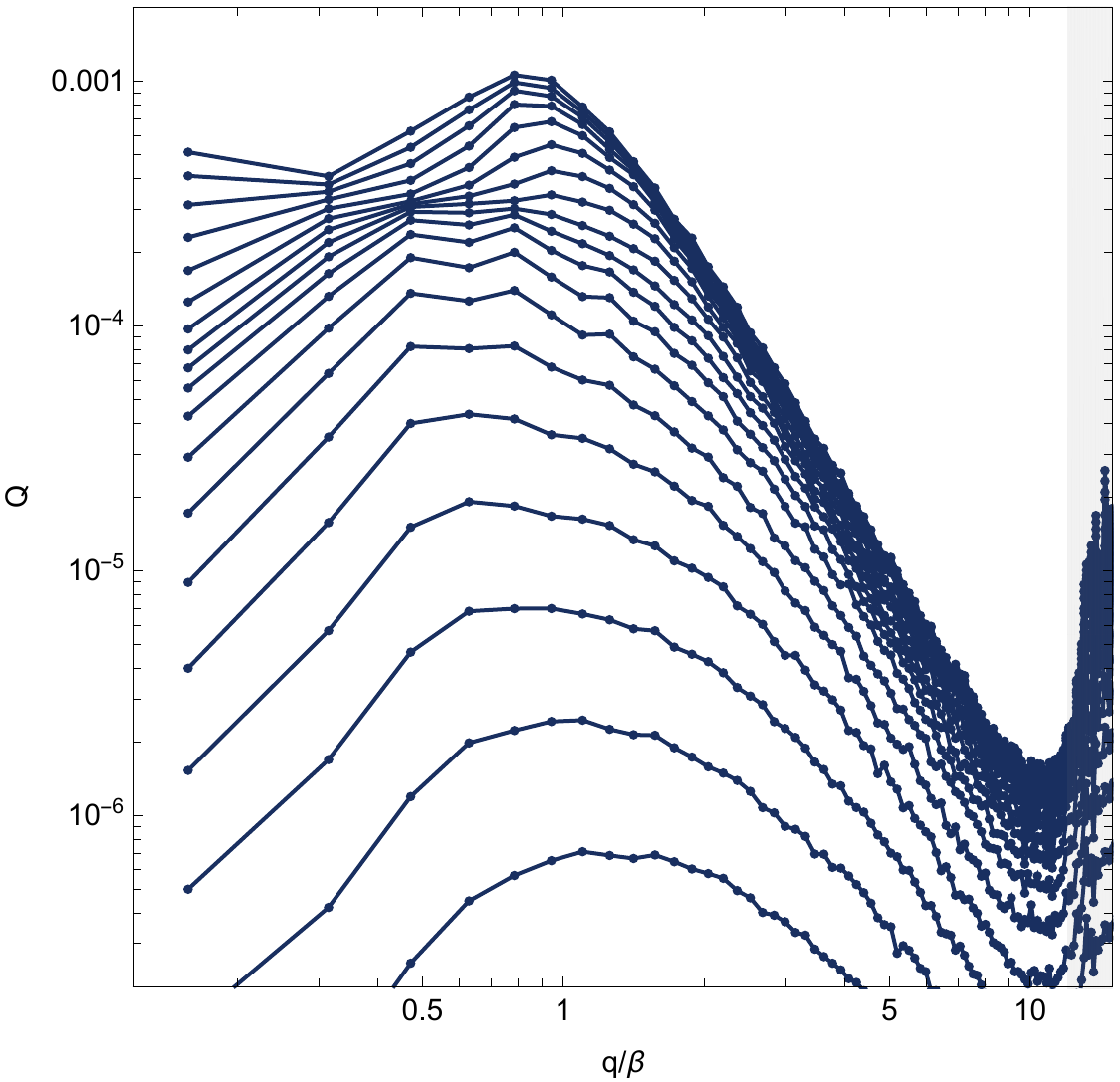}
\includegraphics[width=0.3\textwidth]{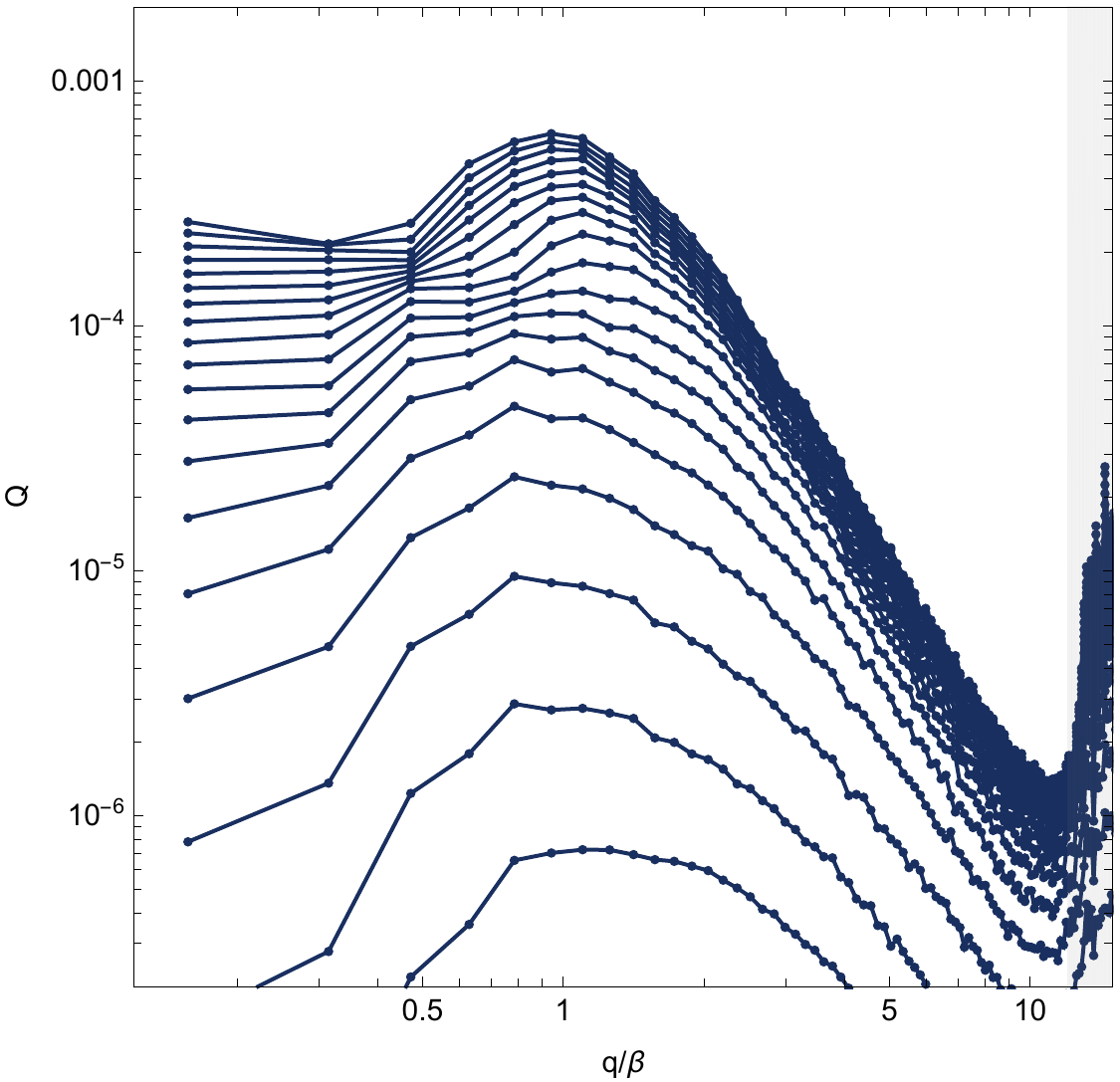} \\
\includegraphics[width=0.3\textwidth]{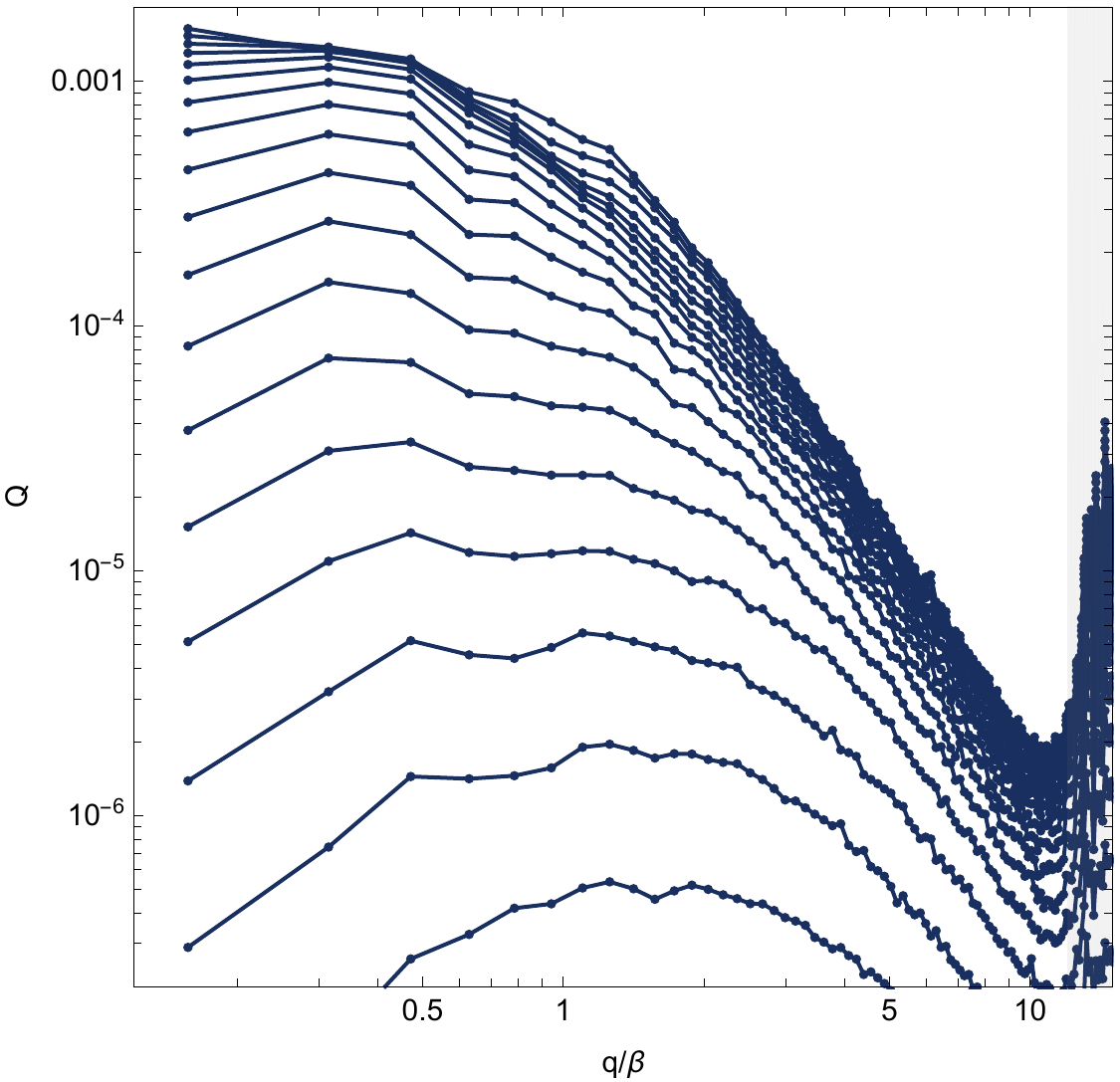}
\includegraphics[width=0.3\textwidth]{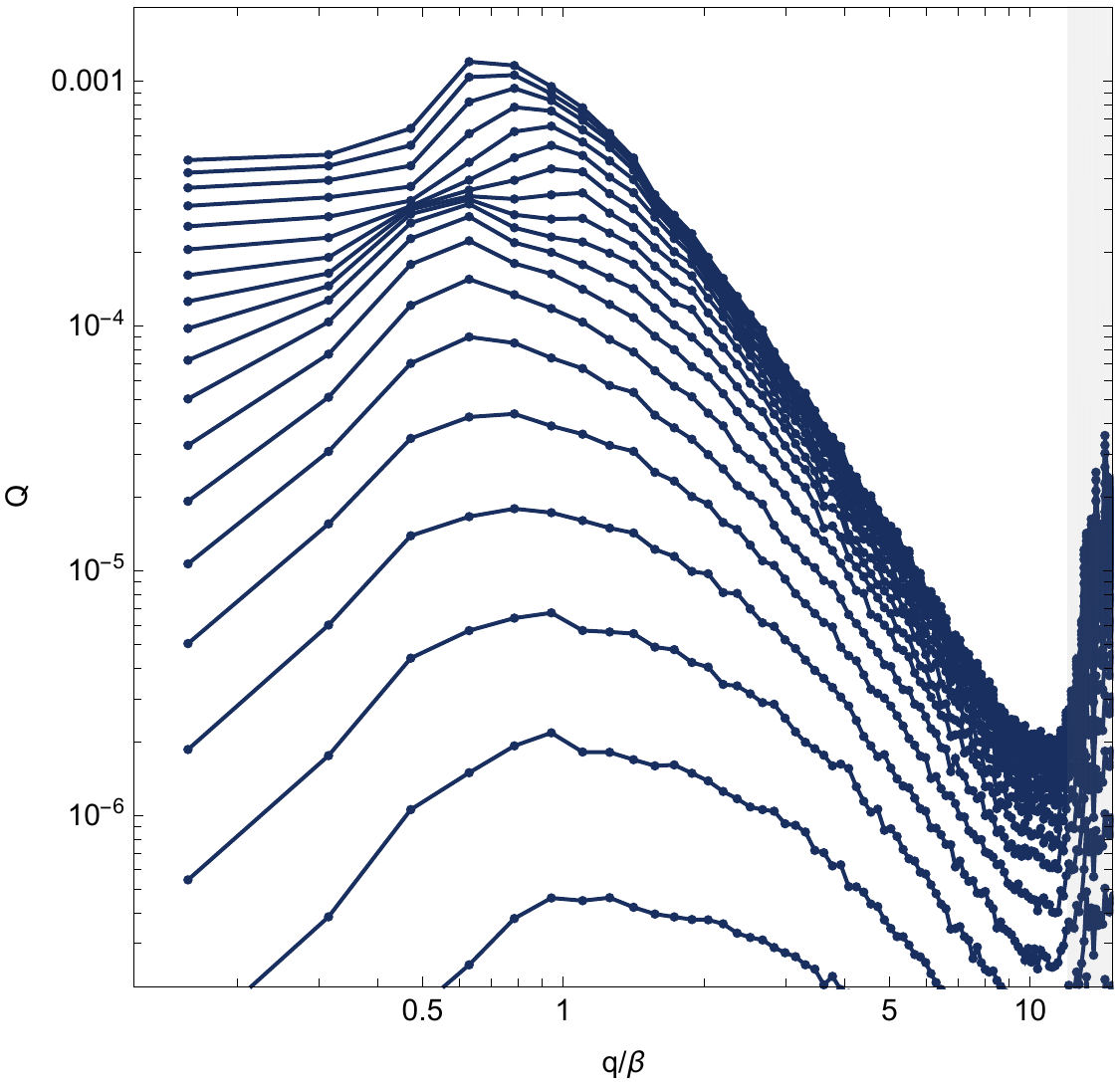}
\includegraphics[width=0.3\textwidth]{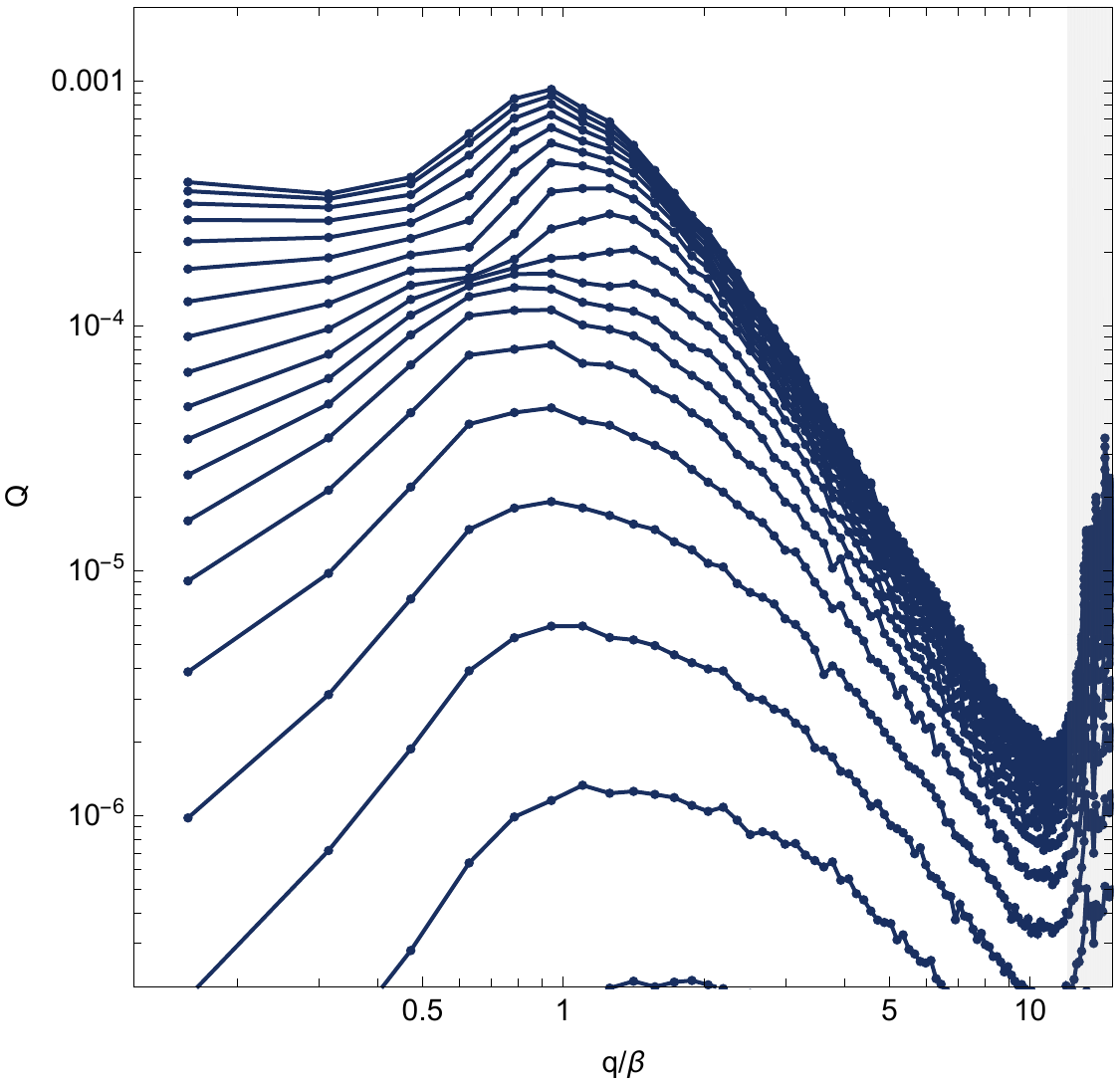}
\caption{
Gravitational-wave spectrum with temperature fluctuations with $c_s = 0$ (top) and $c^2_s = 1/3$ (bottom).
We used the (normalized) temperature fluctuation $\sigma = 3$, and changed the typical scale of the fluctuation $k_* = 2, 4, 6 \times (2\pi/L)$ (left, center and right, respectively).
Otherwise the parameter choices and time slices are the same as in Fig.~\ref{fig:OmegaGW_0}.
}
\label{fig:OmegaGW}
\end{figure}
%%%%%%%%%%

%%%%%%%%%%
\begin{figure}
\centering
\includegraphics[width=0.32\textwidth]{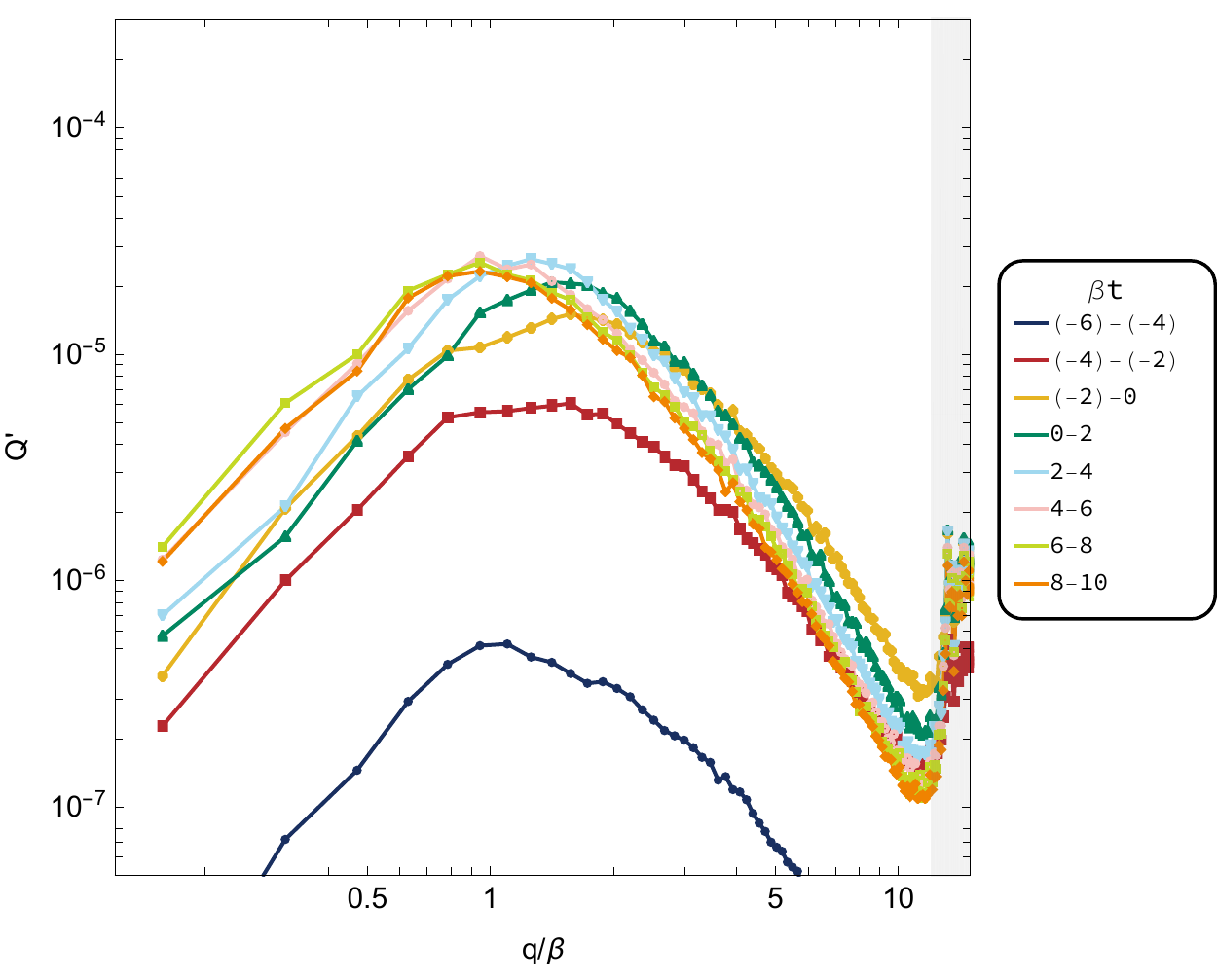}
\caption{
GW emission per unit time $Q'$ calculated between each short time interval $\Delta t = 2 / \beta$.
This figure corresponds to Fig.~\ref{fig:OmegaGW_0}.
}
\label{fig:OmegaGW_0_slice}
\vskip 1cm
\includegraphics[width=0.32\textwidth]{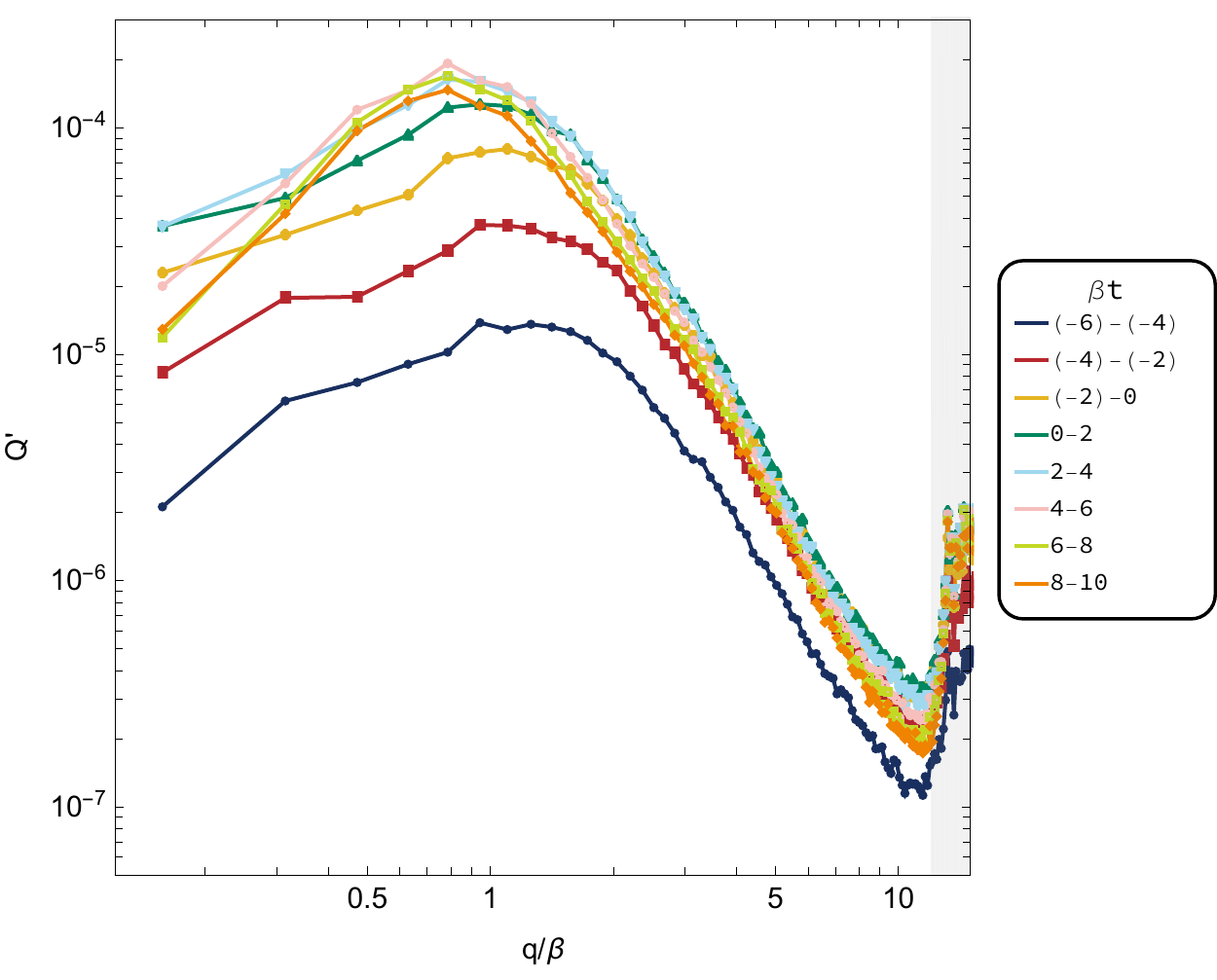}
\includegraphics[width=0.32\textwidth]{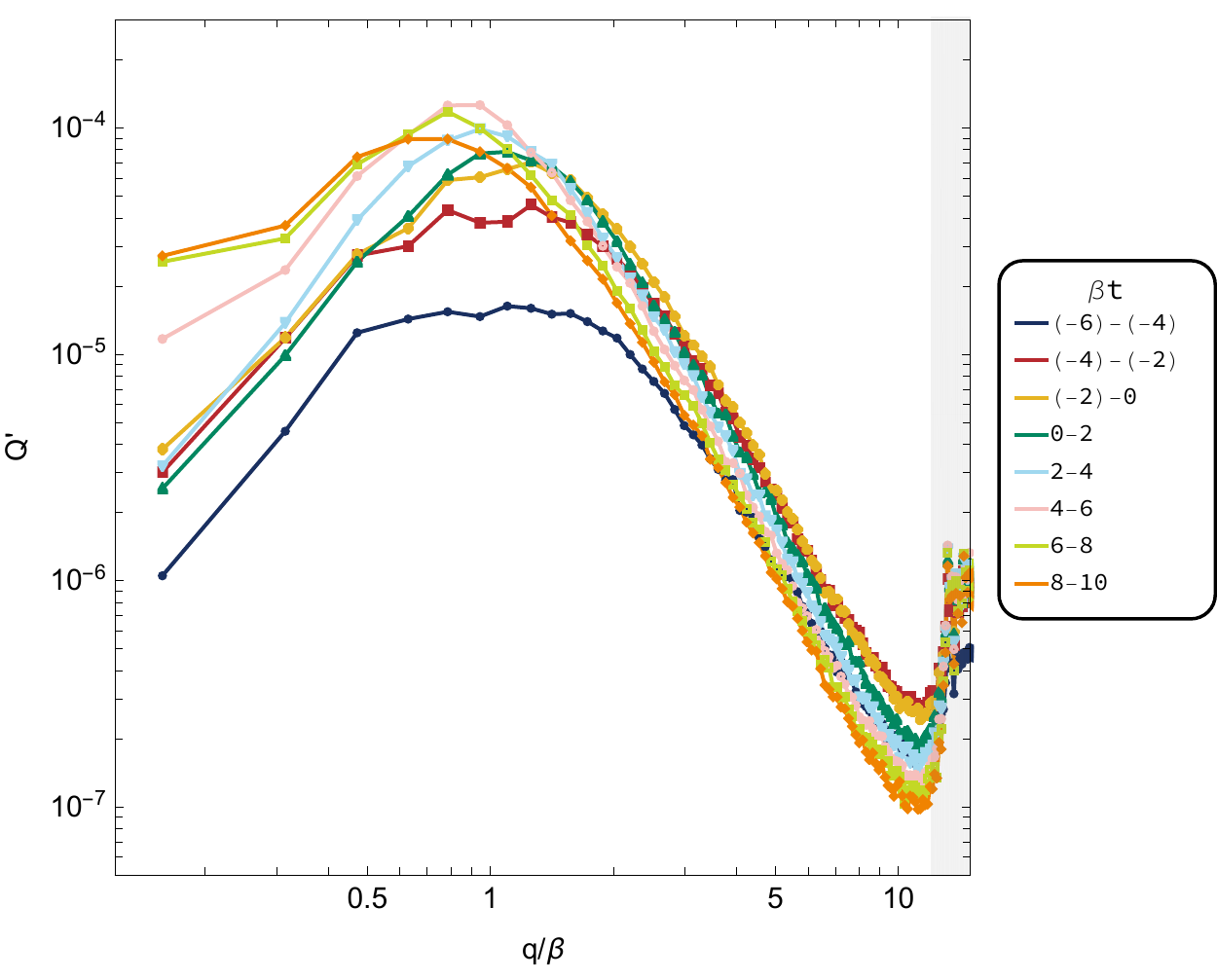}
\includegraphics[width=0.32\textwidth]{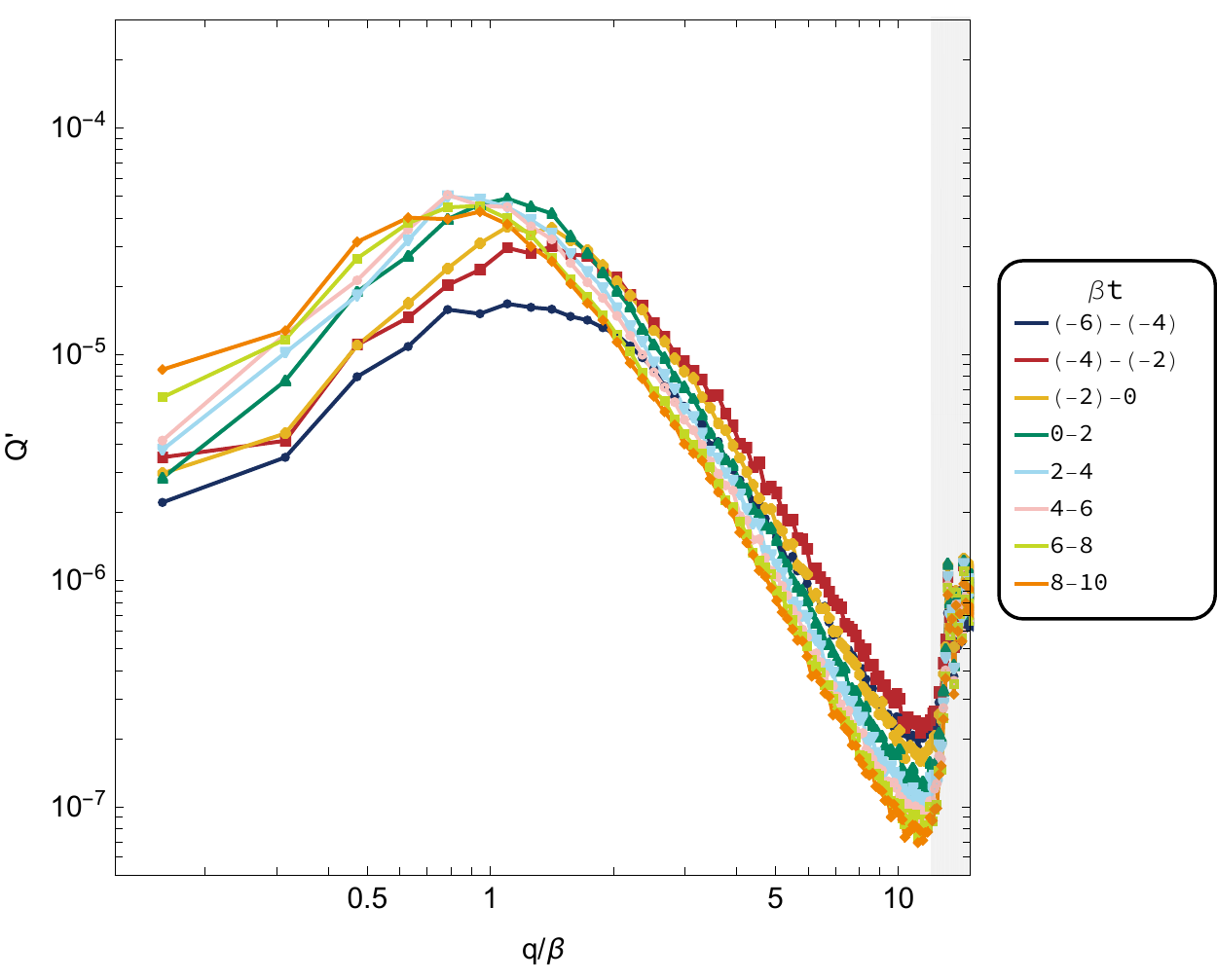}\\
\includegraphics[width=0.32\textwidth]{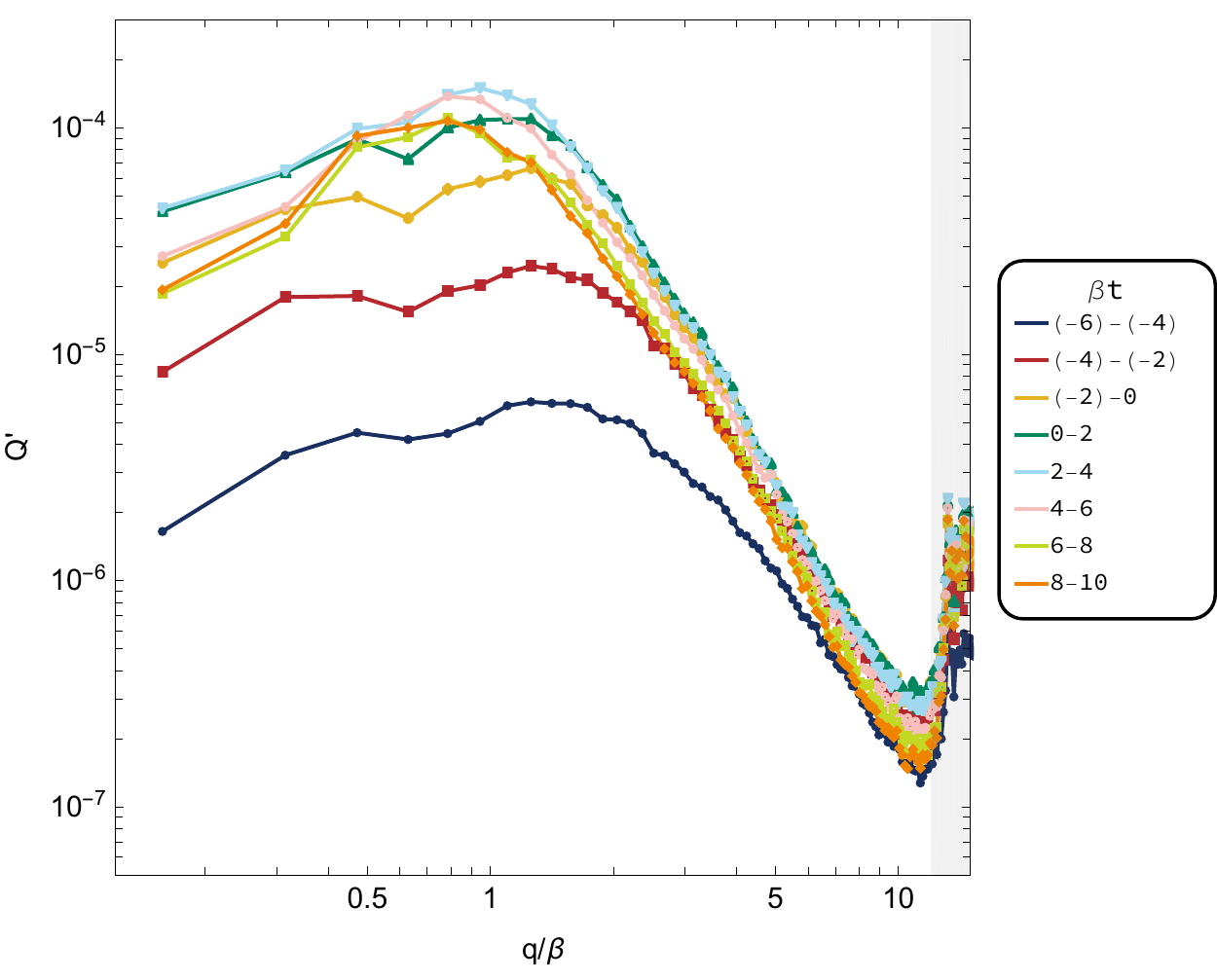}
\includegraphics[width=0.32\textwidth]{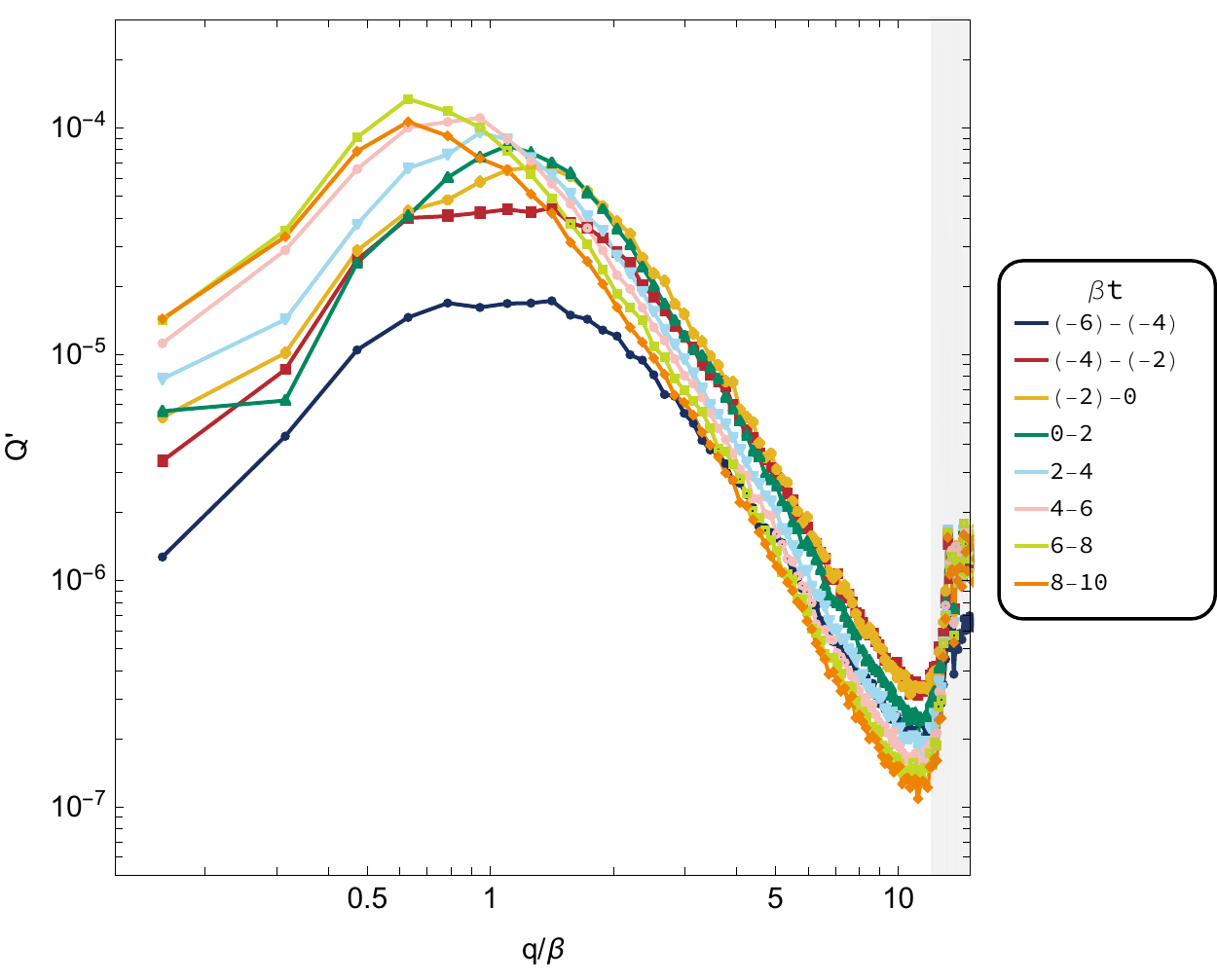}
\includegraphics[width=0.32\textwidth]{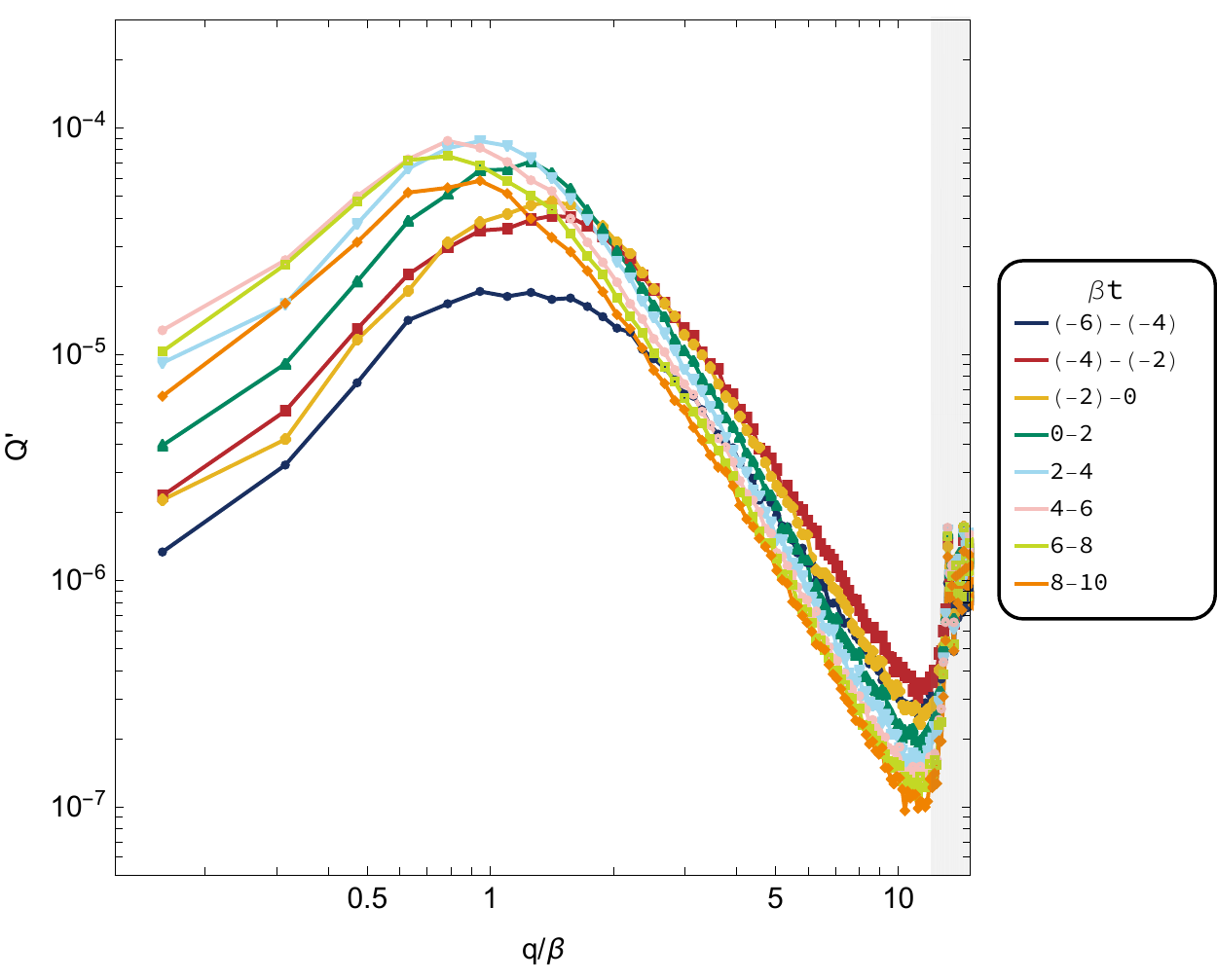}
\caption{
GW emission per unit time $Q'$ calculated between each short time interval $\Delta t = 2 / \beta$ with $c_s = 0$ (top) and $c^2_s = 1/3$ (bottom).
This figure corresponds to Fig.~\ref{fig:OmegaGW}.
}
\label{fig:OmegaGW_slice}
\end{figure}
%%%%%%%%%%
Even though bubble nucleation statistics and histories give some insight on 
the impact of the temperature fluctuations, the actual observable
we are interested in is the GW spectrum generated by the phase transition. 
We next show the numerical results for the GW spectrum.
The benchmark point we use is $L = 40 / \beta$ and wall velocity $v_w = 1$.\footnote{ We obtained similar results for several values of $v_w$. As long as the wall velocity is not much smaller than $c_s$, we expect qualitatively similar effects in the GW enhancement. 
}
We show the GW spectrum without the temperature fluctuations in Fig.~\ref{fig:OmegaGW_0}.
The different lines correspond to different time slices from $t = -10 / \beta$ to $10 / \beta$ with $\Delta t = 1 / \beta$.

In the top panels of Fig.~\ref{fig:OmegaGW} we show the GW spectrum for static temperature fluctuations with ($c_s = 0$).
We used the (normalized) temperature fluctuation $\sigma = 3$, and varied the typical scale of the fluctuation $k_* = 2,4,6 \times (2\pi/L)$ (left, center and right respectively).
In the bottom panels of the same figure, we show the GW spectrum with the temperature fluctuations with $c_s^2 =1/3$ for the same values of $\sigma$ and $k_*$. In both cases, we observe that the IR fluctuations enhance the signal in the IR part and also shift the peak towards smaller wavenumbers.
This can be qualitatively understood from the bubble distribution in Fig.~\ref{fig:simulation}. 
However, we would like to reiterate that this is not easily deduced from the total number of bubbles that is nucleated. 
While for static fluctuations ($c_s = 0$) the total number of bubbles is reduced, this is not the case for dynamical fluctuations ($c_s^2 = 1/3$).
One might hence conclude that the effect is less pronounced in the GW spectrum for dynamical fluctuations, which is not true. 
The spatial distributions of the bubble nucleations is essential and, ultimately, one finds a similar enhancement in the 
GW signal for both cases.

As demonstrated in Figs.~\ref{fig:OmegaGW_0} and \ref{fig:OmegaGW}, the GW spectrum integrated from the beginning of the transition ($t \simeq -10 / \beta$) has both IR and UV structures.
The IR structure (see e.g. $q / \beta \simeq 0.5$ in Fig.~\ref{fig:OmegaGW_0}) comes from the typical bubble size around the time of collisions~\cite{Kosowsky:1991ua,Kosowsky:1992vn,Huber:2008hg,Jinno:2016vai,Jinno:2017fby,Konstandin:2017sat}, while the UV structure comes from the thickness of the sound shells~\cite{Hindmarsh:2013xza,Hindmarsh:2015qta,Hindmarsh:2017gnf}.
As predicted by the sound shell model~\cite{Hindmarsh:2016lnk,Hindmarsh:2019phv}, the GW signal from the latter structure grows linearly in time and leads to an enhancement of the GW signal around these wavenumbers.
To extract the latter component below, we examine the GW power at different times of the simulation. 
The results are shown in Figs.~\ref{fig:OmegaGW_0_slice} and \ref{fig:OmegaGW_slice}.
Each of them corresponds to the simulations in Figs.~\ref{fig:OmegaGW_0} and \ref{fig:OmegaGW}, respectively, and these figures show the GW power calculated over each time width $\Delta t = 2 / \beta$ (i.e. $T_{\rm sim} = 2 / \beta$ in Eq.~(\ref{eq:Qpr})).
We see that the GW power has entered a constant regime after $t \simeq 2 / \beta$.

\begin{figure}
\centering
\includegraphics[width=0.45\textwidth]{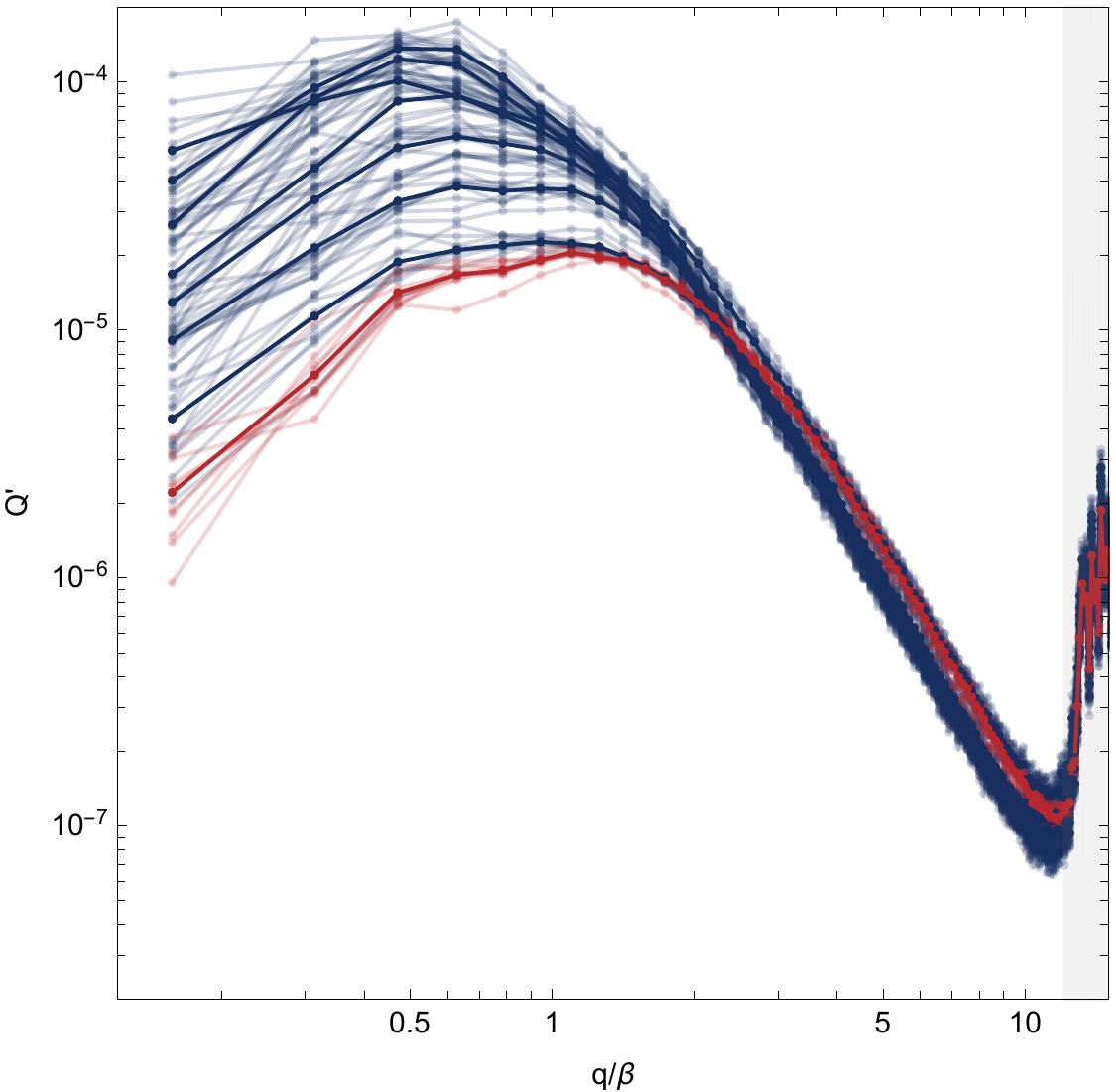}
\vskip 0.3cm
\includegraphics[width=0.45\textwidth]{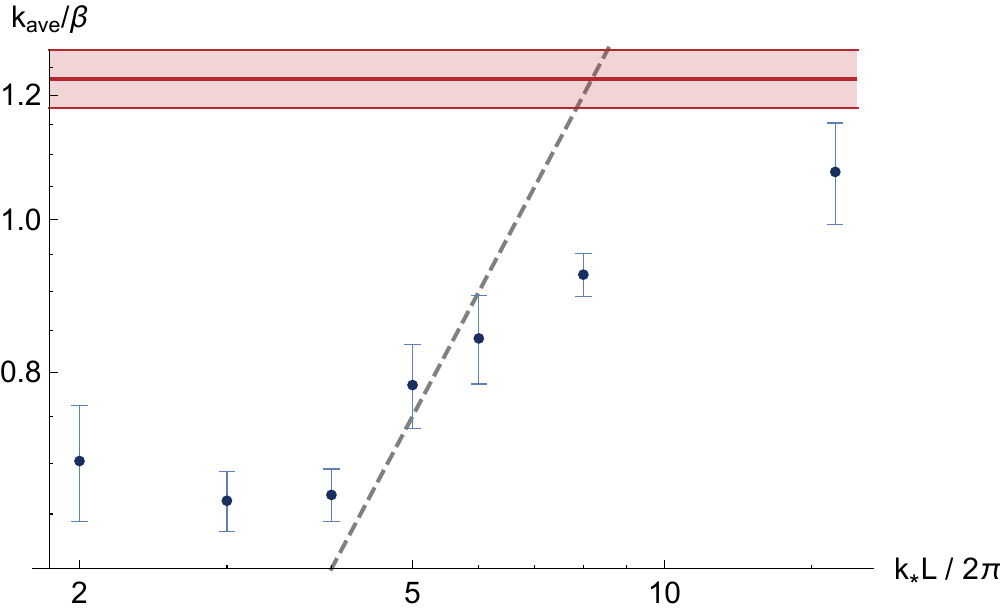}
\includegraphics[width=0.45\textwidth]{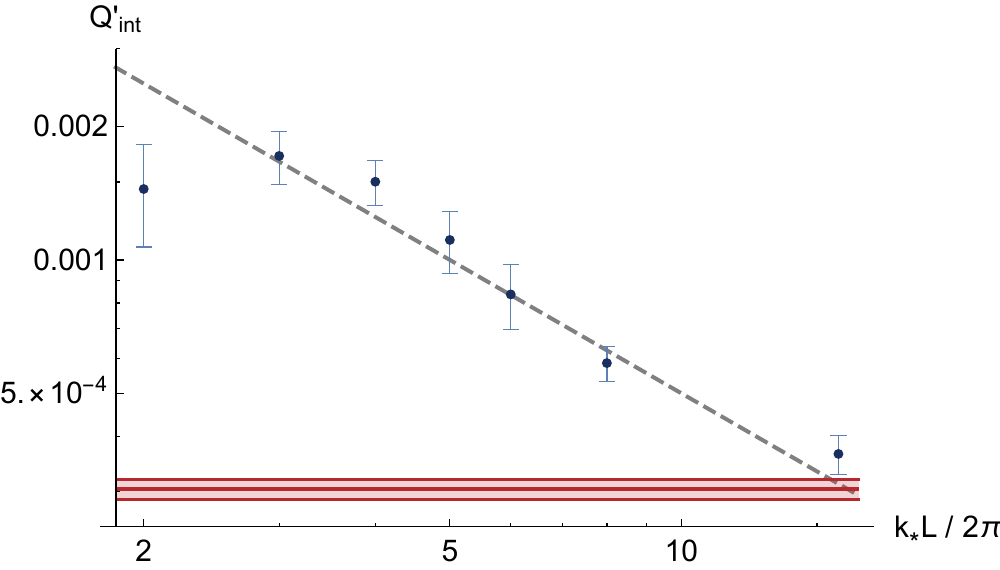}
\caption{
\emph{Top:}
GW power $Q'$ without (red) and with (blue) temperature fluctuation calculated between $2/ \beta < t < 10/ \beta$ with a simulation box of $L = 40 / \beta$.
For the thick blue lines, the fluctuation is set to $\sigma = 3$ with $c_s^2 = 1/3$, and the typical wavenumber is set to $k_* = (2,3,4,5,6,8,16) \times (2 \pi / L)$.
Each thick line is an average of $10$ simulations shown in thin lines.
The thick and thin red lines are the same as the blue ones except that the fluctuation is set to zero.
\emph{Bottom:}
Average GW wavenumber (left) and integrated power (right) for the data shown in the top panel.
The blue data points are for $k_* = (2,3,4,5,6,8,16) \times (2 \pi / L)$, while the red bands are for the case without the temperature fluctuation.
The gray-dashed lines are $\propto k_*$ (left) and $\propto 1 / k_*$ (right) just for comparison.
}
\label{fig:peak}
\end{figure}

Finally, we study the dynamic case in some more detail. We show in the top panel of Fig.~\ref{fig:peak} the GW power $Q'$ without (red) and with (blue) temperature fluctuation.
The power is calculated over $2/ \beta < t < 10/ \beta$ with a simulation box of $L = 40 / \beta$, and the fluctuation is set to $\sigma = 3$ with $c_s^2 = 1/3$.
The typical wavenumber is set to $k_* = (2,3,4,5,6,8,16) \times (2 \pi / L)$ for each thick blue line, which is an average of $10$ simulations shown with thin lines.
The red lines are the same as the blue ones except that the fluctuation is set to zero.
As discussed above, the choice $2/ \beta < t < 10/ \beta$ makes the data free of the IR structure.
The bottom panels of Fig.~\ref{fig:peak} are the average wavenumber (left panel) and the integrated power (right panel) extracted from the data in the top panel.
They are defined as
\begin{align}
k_{\rm ave}
&\equiv
\int d\ln k~
k~Q' (k)
\Big/
\int d\ln k~
Q' (k),
\\
Q_{\rm int}'
&\equiv
\int d\ln k~
Q' (k).
\end{align}
We see that the average wavenumber, as well as the integrated power, show the trends mentioned before: the fluctuations enhance the GW signal and shift it to smaller wavenumbers. 
The data shows that the amplitude approximately scales with $\beta/k_*$ while the scaling of $k_{\rm ave}$ is not obvious.
This is due to the fact that for fluctuations with a small wavenumber ($k_* \sim {\rm a~few} \times (2 \pi / L)$) the simulation volume limits the accuracy. For large wavenumbers, the GW spectrum does not just display a single peak but it has a richer structure coming from the bubble sizes and sound shell thickness~\cite{Jinno:2020eqg}.

%%%%%%%%%%
%%%%%%%%%%

%%%%%%%%%%%%%%%%%%%%%%%%%%%%%%%%%%%%%%%%%%%%%%%%%%
\section{Conclusions}
\label{sec:DC}
%%%%%%%%%%%%%%%%%%%%%%%%%%%%%%%%%%%%%%%%%%%%%%%%%%

In this paper we point out the possible enhancement of the GW signal in first-order phase transitions due to density fluctuation at small scales.
In general, fluctuations in the temperature become important when
\be
\frac{\delta T}{\overline{T}} \gtrsim \frac{H_*}{\beta}.
\ee
This constraint is fulfilled already for rather small fluctuations since for a typical phase transition one often finds values of $\beta/H_*$ larger than a few 100s.

We show that for UV fluctuations (where the typical scale of the fluctuation is larger than the bubble size at the collision time) the effect reduces to an average time shift, and it leaves no net effect on the GW signal. On the other hand, for IR fluctuations we point out that the resulting GW signal can be much larger than the one without temperature fluctuations.
We consider the cases where the fluctuations are dynamical and behave as waves (with a speed of sound $c_s^2=1/3$) as well as static fluctuations. 
We find considerable differences in the distributions of the nucleated bubbles. However, using hybrid simulations, we conclude that the resulting 
GW spectrum shows a similar enhancement in both cases. Heuristically, the enhancement can be explained by the fact that the effective bubble size is increased in the 
phase transition. This then leads to an enhancement in the amplitude as well as a shift of the spectrum to smaller wavenumbers.

%%%%%%%%%%%%%%%%%%%%%%%%%%%%%%%%%%%%%%%%%%%%%%%%%%
\section*{Acknowledgment}
%%%%%%%%%%%%%%%%%%%%%%%%%%%%%%%%%%%%%%%%%%%%%%%%%%

The work of RJ was supported by Grants-in-Aid for JSPS Overseas Research Fellow (No. 201960698).
This work is supported by the Deutsche Forschungsgemeinschaft 
under Germany's Excellence Strategy -- EXC 2121 ,,Quantum Universe`` -- 390833306.

\appendix

%%%%%%%%%%%%%%%%%%%%%%%%%%%%%%%%%%%%%%%%%%%%%%%%%%
\section{Details of the nucleation algorithm}
\label{app:algorithm}
%%%%%%%%%%%%%%%%%%%%%%%%%%%%%%%%%%%%%%%%%%%%%%%%%%

In this appendix we briefly describe the nucleation algorithm used in the main text.
Note that for the GW simulation algorithm we use the calculation scheme proposed in Ref.~\cite{Jinno:2020eqg}.

\begin{figure}
\centering
\includegraphics[width=0.7\textwidth]{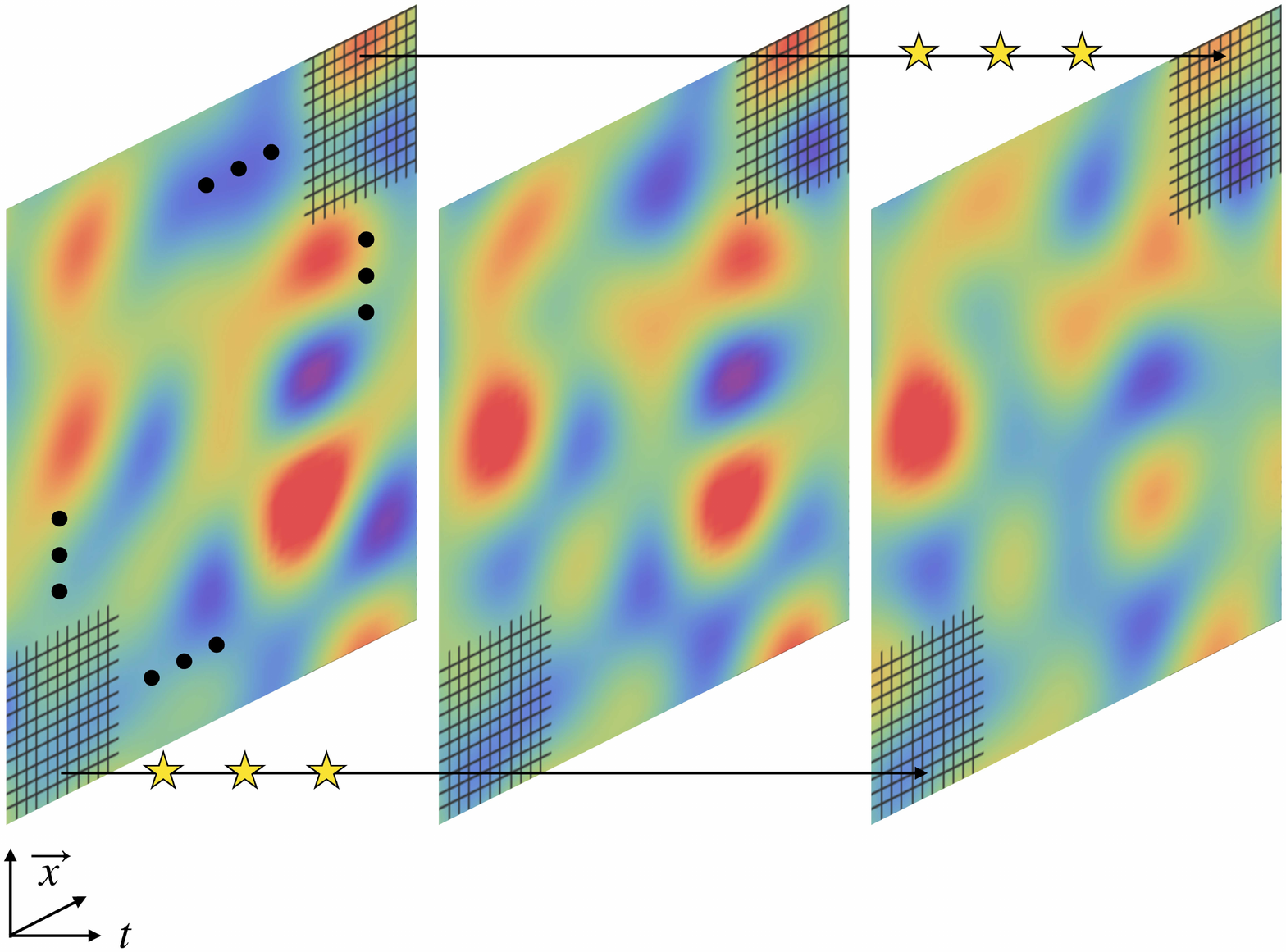} 
\caption{
Illustration for the bubble nucleation algorithm.
}
\label{fig:algorithm}
\vskip 1cm
\includegraphics[width=\textwidth]{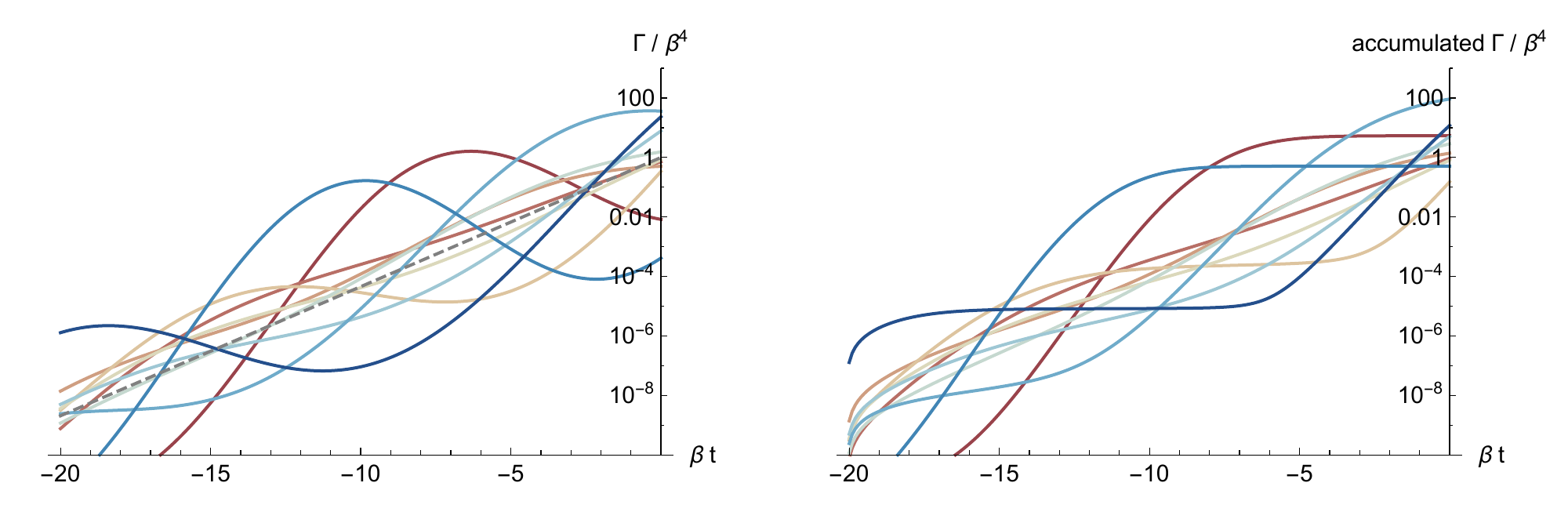} 
\caption{
\emph{Left:} Nucleation rate per unit time and volume in $\beta$ unit for $10$ randomly chosen cells in a simulation with $L = 40 / \beta$, $k_* = 4 \times (2\pi/L)$, and $\sigma = 3$.
The gray-dashed line is the nucleation rate without the temperature fluctuations.
\emph{Right:} Accumulated probability for the $10$ cells in the left panel.
}
\label{fig:prob}
\end{figure}

The bubble nucleation algorithm is demonstrated in Fig.~\ref{fig:algorithm}.
We generate the density fluctuations by randomly sampling $40$ modes satisfying $k_* / 2 \leq k \leq k_*$.
We divide the simulation box $L^3$ into small cells (with $\Delta L = 1 / \beta$), and for each cell we calculate the accumulated nucleation probability in the presence of the density fluctuation as illustrated in Fig.~\ref{fig:prob}. 
The nucleation time is obtained from random seeds sampled from the range of the accumulated probability. 
In practice we calculate the time when the expected number of bubbles reaches $10$ for each cell, and generate $10$ bubbles from $t = -\infty$ up to this point.
For each bubble nucleated we assign a random spatial position within each cell.
Typically, the first bubble nucleated in each cell will cover the cell quite quickly and the remaining nucleation points are discarded since the first nucleated bubble lies in their past light cone.

%%%%%%%%%%
%%%%%%%%%%

%%%%%%%%%%%%%%%%%%%%%%%%%%%%%%%%%%%%%%%%%%%%%%%%%%
\section{Analytic expressions}
\label{app:analytic}
%%%%%%%%%%%%%%%%%%%%%%%%%%%%%%%%%%%%%%%%%%%%%%%%%%

In this section we derive analytic expressions for the bubble nucleation distributions that can be useful when interpreting the results in the main sections.

%%%%%%%%%%%%%%%%%%%%%%%%%%%%%%%%%%%%%%%%%%%%%%%%%%
\paragraph{Without temperature fluctuation}
%%%%%%%%%%%%%%%%%%%%%%%%%%%%%%%%%%%%%%%%%%%%%%%%%%

\begin{figure}[t]
\centering
\includegraphics[width=0.4\textwidth]{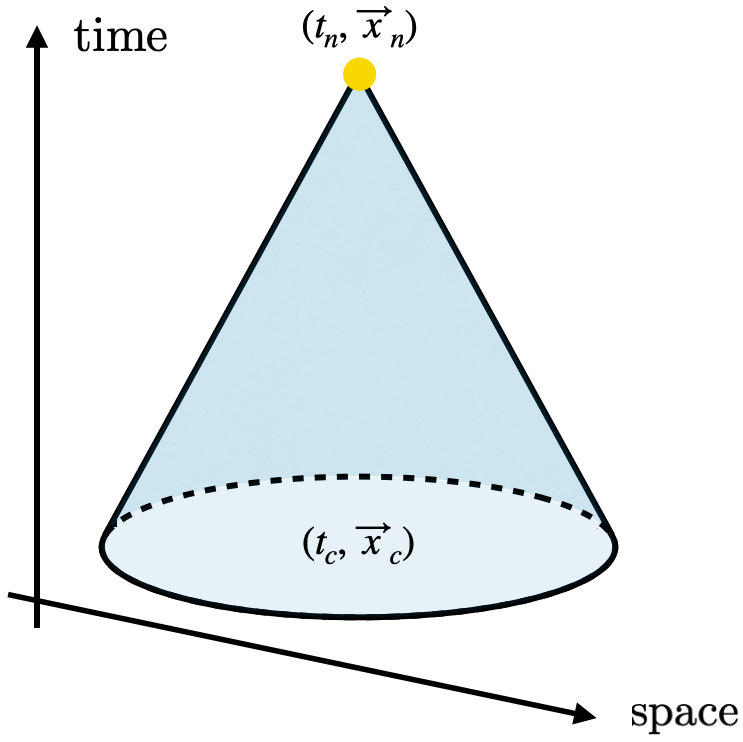}
\caption{
Parametrization for the nucleation time distribution.
}
\label{fig:cone}
\end{figure}

We derive the distribution of the bubble nucleation time.
We first neglect the effect of the temperature fluctuation.
In this case, the nucleation rate can be written simply as
\begin{align}
\Gamma (t, \vec{x})
&= 
\Gamma (t)
= 
\Gamma_* e^{\beta (t - t_*)}.
\end{align}
Here we choose the time $t_*$ so that $\Gamma (t_*) = \Gamma_* = \beta^4$ is satisfied, and label this time as $t_* = 0$.
Though we assume $v_w = 1$ in the following, generalization to $v_w \neq 1$ is straightforward.
For a bubble to nucleate at a given four-dimensional point $(t_n, \vec{x}_n)$ with an infinitesimal spacetime volume element $d^4 x_n = d t_n d^3 x_n$, we need two conditions (see Fig.~\ref{fig:cone}):
\begin{itemize}
\item
No bubble nucleates inside the past light cone of $(t_n, \vec{x}_n)$.
\item
One bubble nucleates in $d^4 x_n$.
\end{itemize}
The former probability, which we call survival probability $P_{\rm surv} (t_n, \vec{x}_n)$, can be expressed as
\begin{align}
P_{\rm surv} (t_n, \vec{x}_n)
&= 
\prod_{x_c \; \in {\rm \; past \; light \; cone \; of \;} (t_n, \vec{x}_n)} \left[
1 - \Gamma (x_c) \; d^4 x_c
\right]
\nonumber \\
&=
\exp \left[
- \int_{x_c \; \in {\rm \; past \; light \; cone \; of \;} (t_n, \vec{x}_n)} d^4 x_c~\Gamma (x_c)
\right]
\nonumber \\
&=
e^{- 8 \pi e^{\beta t_n}}.
\end{align}
In the last equality we neglected the effect of cosmic expansion.
Together with the latter probability $\propto e^{\beta t_n}$, the nucleation time distribution $P_{n, \delta \tilde{T} = 0} (t_n)$ is
\begin{align}
P_{n, \delta \tilde{T} = 0} (t_n)
&= 
8 \pi e^{\beta t_n - 8 \pi e^{\beta t_n}}.
\end{align}
This is plotted as the red lines in Figs.~\ref{fig:tNuc_indep} and \ref{fig:tNuc_dep}.
Note that the overall factor is chosen so that $\int_{- \infty}^\infty d t_n \; P_{n, \delta \tilde{T} = 0} (t_n) = 1$.
Note also that the expression before normalizing with $8 \pi$ gives the average number of bubbles $N_b = (\beta L)^3 / 8 \pi$.

%%%%%%%%%%%%%%%%%%%%%%%%%%%%%%%%%%%%%%%%%%%%%%%%%%
\paragraph{With temperature fluctuation}
%%%%%%%%%%%%%%%%%%%%%%%%%%%%%%%%%%%%%%%%%%%%%%%%%%

Next we include the temperature fluctuation.
While it is difficult to evaluate the final expression analytically, we can extract the IR and UV limits from it.
The nucleation rate is
\begin{align}
\Gamma (t, \vec{x})
&= 
\Gamma_* e^{\beta (t - t_*) - \delta \tilde{T} (t, \vec{x})}.
\end{align}
We again take $\Gamma_* = \beta^4$ and $t_* = 0$.
The survival probability becomes (see Fig.~\ref{fig:cone})
\begin{align}
P_{\rm surv} (t_n, \vec{x}_n)
&=
\exp \left[
- \int_{x_c \; \in {\rm \; past \; light \; cone \; of \;} (t_n, \vec{x}_n)} d^4 x_c~\Gamma (x_c)
\right],
\end{align}
and thus the nucleation time distribution $P_n (t_n)$ averaged over all possible temperature configurations is given by
\begin{align}
P_n (t_n)
&= 
N
\left< 
\Gamma (t_n, \vec{x}_n)
\exp \left[
- \int_{x_c \; \in {\rm \; past \; light \; cone \; of \;} (t_n, \vec{x}_n)} d^4 x_c~\Gamma (x_c)
\right]
\right>_{\rm ens}.
\label{eq:Pn_with_T}
\end{align}
Here $\left< \cdots \right>_{\rm ens}$ is the ensemble average over the temperature fluctuation.
The normalization factor $N$ should be chosen so that the total probability becomes unity.

%%%%%%%%%%
%%%%%%%%%%

%%%%%%%%%%%%%%%%%%%%%%%%%%%%%%%%%%%%%%%%%%%%%%%%%%
\paragraph{IR and UV limits}
%%%%%%%%%%%%%%%%%%%%%%%%%%%%%%%%%%%%%%%%%%%%%%%%%%

We consider the IR limit $k_* \to 0$ and the UV limit $k_* \to \infty$ with fixed $\sigma$.
In the IR limit, the temperature is the same over scales much larger than the typical separation between bubbles.
Assuming that the temperature fluctuation is Gaussian, and noting that $\delta \tilde{T}$ is essentially a time shift, we obtain
\begin{align}
P_{n, {\rm IR}} (t_n)
&= 
\frac{1}{\sqrt{2 \pi \sigma^2}}
\int d (\beta \delta t_n)~
e^{- \frac{(\beta \delta t_n)^2}{2 \sigma^2}}
P_{n, \delta \tilde{T} = 0} (t_n + \delta t_n).
\label{eq:Pn_IR}
\end{align}
We evaluated the expression (\ref{eq:Pn_IR}) which yields the red-dotted lines in Figs.~\ref{fig:tNuc_indep} and \ref{fig:tNuc_dep}.

In the UV limit, the two factors inside the ensemble average of Eq.~(\ref{eq:Pn_with_T}) decouples because an infinite number of temperature fluctuation is realized inside the past cone of $(t_n, \vec{x}_n)$.
More generally, this argument also allows us to simplify the ensemble-averaged survival probability $P_{\rm surv} (t_n)$
\begin{align}
P_{n, {\rm UV}} (t_n)
&= 
N
\left< 
\Gamma (t_n, \vec{x}_n)
\prod_{x_c \; \in {\rm \; past \; light \; cone \; of \;} (t_n, \vec{x}_n)} \left[
1 - \Gamma (x_c) \; d^4 x_c
\right]
\right>_{\rm ens}
\nonumber \\
&=
N
\left< 
\Gamma (t_n, \vec{x}_n)
\right>_{\rm ens}
\prod_{x_c \; \in {\rm \; past \; light \; cone \; of \;} (t_n, \vec{x}_n)} \left[
1 - 
\left< 
\Gamma (x_c)
\right>_{\rm ens}
d^4 x_c
\right].
\end{align}
The ensemble-averaged nucleation rate just gives a time shift
\begin{align}
\left<
\Gamma (t_n, \vec{x}_n)
\right>_{\rm ens}
&=
\frac{\beta^4}{\sqrt{2 \pi \sigma^2}}
\int d (\beta \delta t_n)~
e^{- \frac{(\beta \delta t_n)^2}{2 \sigma^2}}~
e^{\beta (t_n + \delta t_n)}
=
\beta^4 e^{\beta t_n + \frac{\sigma^2}{2}}.
\end{align}
Therefore, we obtain the following UV limit
\begin{align}
P_{n, {\rm UV}} (t_n)
&= 
P_{n, \delta \tilde{T} = 0} \left( t_n + \frac{\sigma^2}{2 \beta} \right).
\end{align}
This is plotted as the red-dashed lines in Figs.~\ref{fig:tNuc_indep} and \ref{fig:tNuc_dep}.

%%%%%%%%%%%%%%%%%%%%%%%%%%%%%%%%%%%%%%%%%%%%%%%%%%
\small
\bibliography{ref}
%%%%%%%%%%%%%%%%%%%%%%%%%%%%%%%%%%%%%%%%%%%%%%%%%%

%%%%%%%%%%%%%%%%%%%%%%%%%%%%%%%%%%%%%%%%%%%%%%%%%%
\end{document}